\documentclass[twocolumn,apj,iop]{openjournal}

\usepackage{graphicx}
\graphicspath{{plots/}}
\usepackage{amsmath}
\usepackage{amssymb}
\usepackage{natbib}
\usepackage{hyperref}
\usepackage{xcolor}
\usepackage{listings}
\usepackage{tikz}
\usepackage{pgfplots}
\usepackage[utf8]{inputenc}
\usepackage{color}
\usepackage[normalem]{ulem} 
\usepackage{physics}
\usepackage{enumitem}
\usepackage{comment}
\usepackage{bm}
\usepackage{aas_macros}
\usepackage{multirow}
\usepackage{xspace}
\hypersetup{
  breaklinks = true,
  colorlinks   = true, 
  urlcolor     = blue!50!black, 
  linkcolor    = blue!50!black, 
  citecolor   = blue!50!black 
}

\usetikzlibrary{arrows.meta, positioning, fit, backgrounds, calc}

\lstdefinestyle{pythonstyle}{
  language=Python,
  basicstyle=\ttfamily\small,
  keywordstyle=\color{blue!60!black}\bfseries,
  commentstyle=\color{gray},
  stringstyle=\color{purple!60!black},
  showstringspaces=false,
  breaklines=true,
  columns=fullflexible,
  frame=lines,
  framesep=3pt,
  xleftmargin=3pt,
  xrightmargin=3pt,
}

\newcommand{\justifycaption}{\leftskip 0pt\rightskip 0pt\parfillskip 0pt plus 1fil\relax}

\IfFileExists{fontawesome5.sty}{
  \usepackage{fontawesome5}
}{
  \newcommand{\faGithub}{[GitHub]}
  \newcommand{\faOrcid}{[ORCID]}
}
\definecolor{orcidlogocol}{HTML}{A6CE39}
\newcommand{\orcidlink}[1]{\href{https://orcid.org/#1}{\textcolor{orcidlogocol}{\faOrcid}}}
\definecolor{githublogocol}{HTML}{0366D6}
\newcommand{\githublink}[1]{\href{#1}{\textcolor{githublogocol}{\faGithub}}}

\shorttitle{\texttt{bihalofit-JAX}}
\shortauthors{Alexander Roskill \& Matteo Zennaro}

\begin{document}

\title{\texttt{bihalofit-JAX}: A Differentiable Code for the
Nonlinear Matter Bispectrum and its Projected Statistics\vspace{-4em}}
\author{Alexander Roskill\altaffilmark{1,$\star$}\,\orcidlink{0009-0009-8861-1618} and Matteo Zennaro\altaffilmark{2}\,\orcidlink{0000-0002-4458-1754}}
\affiliation{\altaffilmark{1}Astrophysics, University of Oxford, Denys Wilkinson Building, Keble Road, Oxford OX1 3RH, United Kingdom\\
\altaffilmark{2}Institute of Space Sciences (ICE, CSIC), Campus UAB,
Carrer de Can Magrans, s/n, 08193 Barcelona, Spain}
\email{$\star$alexander.roskill@physics.ox.ac.uk}

\begin{abstract}
Theoretical predictions for the nonlinear matter bispectrum and related projected statistics have been historically very expensive to evaluate, leading many analyses to rely on emulators to speed up slow numerical codes. We show in this paper that this need not be the case: improved numerical techniques can be leveraged to directly evaluate predictions for the nonlinear matter bispectrum and projected statistics at speeds preferable for inference. We present \texttt{bihalofit-JAX}, an open-source, fully differentiable code for the nonlinear matter bispectrum based on the BiHalofit fitting formula, which includes numerous related projected statistics such as angular bispectra, skew spectra and aperture statistics. It computes $10000$ triangles of the matter bispectrum in $\sim1$--$2\ \rm{ms}$ per cosmology, one to three orders of magnitude faster than other implementations. It agrees numerically with the original BiHalofit code to better than $0.02\%$ on average over the cosmologies the formula was calibrated on, where the formula itself is accurate to $\sim20\%$ against simulations. It is also efficiently vectorised such that one can compute e.g. $500$ cosmologies in $\sim 2\ \rm{ms}$. This speed-up grows even more in the case of the third-order aperture statistics, which we compute in $\sim 6\ \rm{ms}$ for one cosmology, between $3\times10^3$ and $2.4\times10^4$ times faster than other publicly available codes. The pipeline is fully differentiable, and we demonstrate its performance in a simple inference problem, comparing the No-U-Turn Sampler (NUTS) to gradient-free inference with the affine-invariant ensemble sampler (\texttt{emcee}). \githublink{https://github.com/AlexRoskill/bihalofit-jax}
\end{abstract}

\section{Introduction}

The matter bispectrum (the three-point correlation function of the density field in Fourier space) contains information that is inaccessible to the power spectrum alone \citep{Scoccimarro_2000}. Because it is sensitive to the non-Gaussian structure generated by gravitational collapse, it helps to constrain bias parameters \citep{Philcox_2022, Harscouet2026Constraints} and to break parameter degeneracies \citep{bernardeau1997weaklensingstatisticsprobe,Takada_2004,Sefusatti_2006}. It is also the primary large scale structure probe of primordial non-Gaussianity, whose signature is imprinted in the initial conditions \citep{Cabass_2022, D_Amico_2025,cabass2025bossconstraintsmassiveparticles,chudaykin2025reanalyzingdesidr13}. Exploiting this information in forthcoming galaxy and lensing surveys, however, requires techniques that have lagged well behind those for the power spectrum (two-point correlation function in Fourier space): measurements are costly \citep{xie2026efficientestimatorspowerspectrum}, covariance matrices need large numbers of mocks \citep{Philcox_2021}, and theoretical predictions are both expensive \citep{Chudaykin_2020,Chen:2024pyp} and difficult to model accurately \citep{McCullagh_2015}. 

BiHalofit \citep{Takahashi2020} is a widely used analytic fitting formula for the nonlinear
matter bispectrum, calibrated directly against $N$-body simulations. It achieves an accuracy of $\sim 20\%$ on scales of $k< 3 \ h \ \rm{Mpc}^{-1}$ and redshifts of  $z=0$--$1.5$ for $w$-cold dark matter ($w$CDM) cosmologies ($w$ refers to a constant dark energy equation of state). It
includes a prescription for correcting for baryonic effects, which are critically important on small scales \citep{Foreman_2020,Aric__2021,Burger_2026}. BiHalofit offers a powerful alternative
to perturbation-theory approaches, which break down on mildly
nonlinear scales and neglect information contained in the nonlinear regime. Existing public implementations of this model, however, have historically been extremely slow to evaluate, particularly for projected statistics such as the weak-lensing convergence bispectrum and mass-aperture statistics \citep{Kayo_2012,Heydenreich2023}, which require multiple nested integrals over the matter bispectrum. The reason for this is that the codes available for BiHalofit and its related projected statistics are not sufficiently optimised (particularly for modern hardware e.g.\ GPUs), and are not efficiently vectorised (important whenever the statistics must be evaluated over large batches of cosmologies, redshifts, or triangle configurations). Due to these difficulties, many analyses have relied on emulators of statistics related to the bispectrum \citep{Heydenreich2023,Gong_2023, Burger2024,Gomes_2025a,Gomes_2025b, sugiyama2025cosmologyjointanalysissecond,linke2026kidslegacyjointanalysissecond}, which replace computationally expensive codes with fast surrogates trained on evaluations of the codes. Despite their ability to render the inference tractable, emulators introduce additional sources of error \citep{Bevins_2025}, require large training sets \citep{nygaard2024cuttingcornershyperspheresampling}, and are intrinsically limited to the parameter space on which they were trained. As such it is preferable to have a fast, vectorised implementation of the underlying model itself (also ideally differentiable), which can be used directly in inference pipelines without the need for emulation.

\texttt{bihalofit-JAX} is a new implementation of BiHalofit and of the projected statistics built on it: projected bispectra, skew spectra and aperture statistics. Its improved performance compared to existing codes comes from its numerical design: a fixed-iteration Newton solve for the nonlinear scale replaces bisection over an adaptively integrated $\sigma^2(R)$, and fixed quadrature grids, including a change of variables for the aperture statistics, replace adaptive integration. Because every step has a fixed shape and iteration count, the whole calculation compiles once and is batched across cosmologies, redshifts and triangle configurations. We write it in \texttt{JAX} \citep{jax2018github}, which provides the just-in-time (JIT) compilation, GPU portability and automatic differentiation, with no change to the user-facing Python code. The full pipeline is differentiable, so gradients flow from the projected observables back to the cosmological parameters, in the spirit of differentiable pipelines for two-point statistics \citep{Ruiz_Zapatero_2024, Campagne_2023, SwiftCl2025, chiarenza2026blastjldifferentiablenonlimberpower}. This enables gradient-based samplers \citep{Duane:1987de, Brooks_2011,hoffman2011nouturnsampleradaptivelysetting,robnik2026microcanonicalhamiltonianmontecarlo}, which can greatly speed up high-dimensional inference \citep{mootoovaloo2024assessmentgradientbasedsamplersstandard,nygaard2026posteriorsamplingageemulators}.

\begin{figure}[htbp]
\centering
\resizebox{\columnwidth}{!}{%
\begin{tikzpicture}[
  node distance=7mm and 10mm,
  box/.style={
    draw, rounded corners, fill=blue!6,
    align=center, minimum width=48mm, minimum height=9mm,
    inner sep=2mm, font=\small
  },
  backend/.style={
    draw, rounded corners, fill=orange!15,
    align=center, minimum width=24mm, minimum height=8mm,
    inner sep=1.5mm, font=\small
  },
  statbox/.style={
    draw, rounded corners, fill=blue!6,
    align=center, text width=42mm, minimum height=15mm,
    inner sep=1.5mm, font=\small
  },
  arr/.style={-{Latex[length=2mm]}, thick},
  sectag/.style={font=\small, text=gray!70!black}
]
  \node[box] (theta) {Cosmological parameters $\theta$\\
    \footnotesize$(\Omega_{\rm cb},\sigma_8,n_s,h,m_\nu,w_0,w_a)$};
  \node[box, below=of theta] (pl) {\texttt{pk\_backends.py}\\
    \footnotesize Linear power spectrum\\ \footnotesize $P_{\rm L}(k,z)$};
  \node[box, below=of pl] (bhf) {\texttt{bispectrum.py}\\
    \footnotesize BiHalofit kernel ($k_{\rm NL},\,n_{\rm eff}$)\\
    \footnotesize $B_\delta(k_1,k_2,k_3;z)$};
  \node[box, below=of bhf] (proj) {\texttt{projected\_bispectrum.py}\\
    \footnotesize Limber projection\\ e.g. \footnotesize $B_\gamma(\ell_1,\ell_2,\ell_3)$};
  \node[statbox, below=10mm of proj, xshift=-23mm] (skew) {\texttt{skew\_spectra.py}\\
    \footnotesize Skew spectra\\ e.g. \footnotesize $C_\ell^{\gamma\gamma^2}$};
  \node[statbox, below=10mm of proj, xshift=23mm] (map) {\texttt{aperture.py}\\
    \footnotesize Aperture statistics\\  e.g. \footnotesize $\langle M_{\rm ap}^3\rangle$};

  \node[backend] (back1) at ($(pl.west)+(-22mm,11mm)$) {\texttt{baccoemu}};
  \node[backend] (back2) at ($(pl.west)+(-22mm,0mm)$) {\texttt{syren-new}};
  \node[backend] (back3) at ($(pl.west)+(-22mm,-11mm)$) {\texttt{E\&H}};
  \begin{pgfonlayer}{background}
  \node[draw, rounded corners, fill=orange!6, inner sep=4mm,
    label={[font=\small]above:$P_{\rm L}$ backends},
    fit=(back1)(back2)(back3)] (backgroup) {};
  \end{pgfonlayer}

  \draw[arr] (theta) -- (pl);
  \draw[arr] (pl) -- (bhf);
  \draw[arr] (bhf) -- (proj);
  \draw[arr] (proj.south) -- (skew.north);
  \draw[arr] (proj.south) -- (map.north);
  \draw[arr] (backgroup) -- (pl.west);

  \node[sectag, anchor=south east] at ($(pl.south east)+(-0.8mm,0.8mm)$) {\S\ref{sec:backends}};
  \node[sectag, anchor=south east] at ($(bhf.south east)+(-0.8mm,0.8mm)$) {\S\ref{sec:kernel}};
  \node[sectag, anchor=south east] at ($(proj.south east)+(-0.8mm,0.8mm)$) {\S\ref{sec:bkappa}};
  \node[sectag, anchor=south east] at ($(skew.south east)+(-0.8mm,0.8mm)$) {\S\ref{sec:skew}};
  \node[sectag, anchor=south east] at ($(map.south east)+(-0.8mm,0.8mm)$) {\S\ref{sec:mapn}};
\end{tikzpicture}%
}
\caption{Overview of the \texttt{bihalofit-JAX} pipeline. Cosmological
parameters are input by the user, with the linear power spectrum (Section~\ref{sec:backends})
supplied by one of three interchangeable backends. The BiHalofit
kernel then evaluates the nonlinear matter bispectrum, which is
Limber-projected onto an observable (e.g.\ weak-lensing) bispectrum
(Section~\ref{sec:bkappa}) and reduced to downstream summary
statistics: skew spectra (Section~\ref{sec:skew}) and third-order aperture statistics (Section~\ref{sec:mapn}).}
\label{fig:pipeline}
\end{figure}
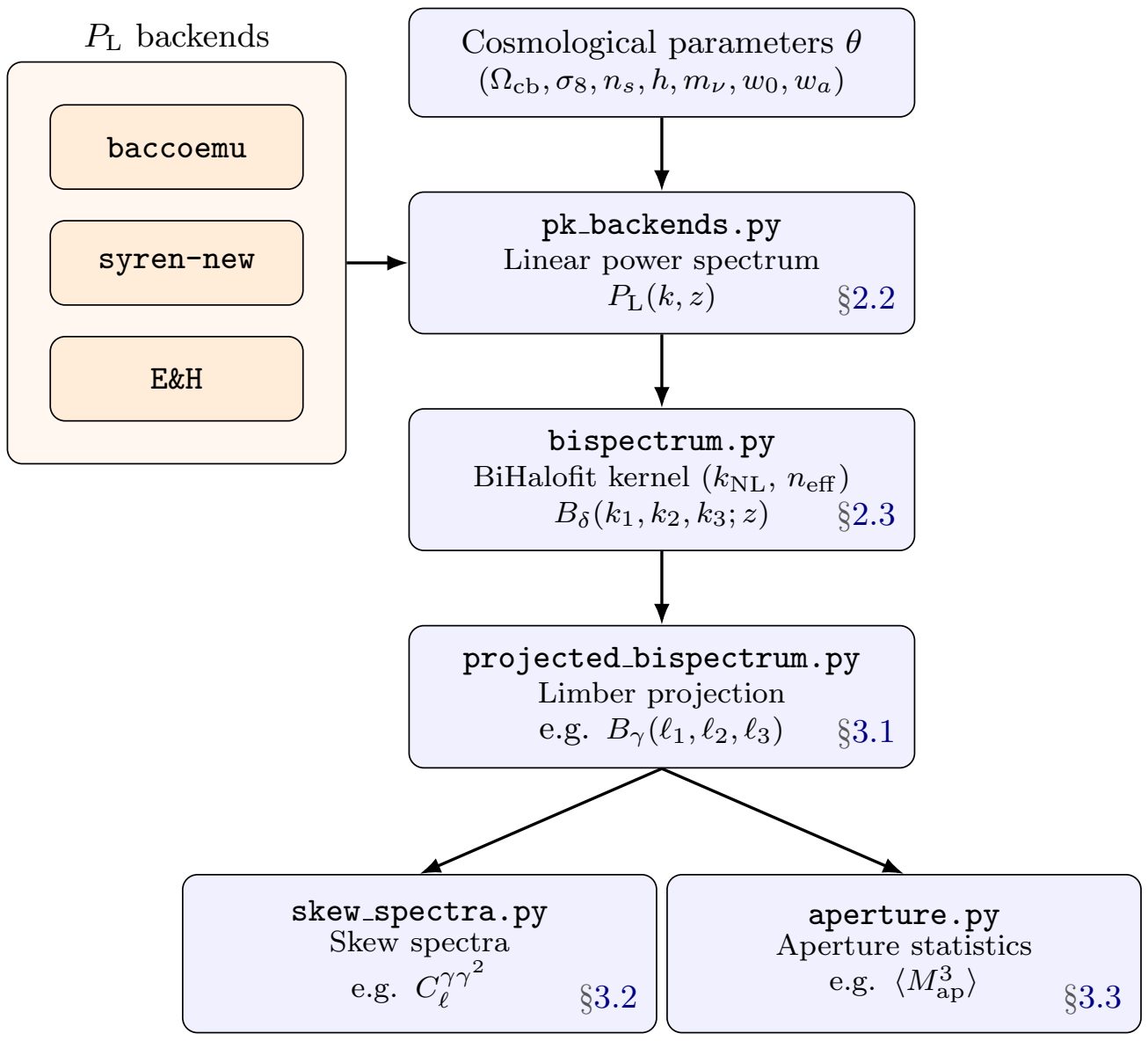

This paper describes the
\texttt{bihalofit-JAX} package, which is summarised in Figure~\ref{fig:pipeline}. We present the main BiHalofit implementation in Section~\ref{sec:model}. We discuss the projected statistics in Section~\ref{sec:projection}: the projected bispectra in Section~\ref{sec:bkappa}, skew spectra in Section~\ref{sec:skew} and aperture statistics in Section~\ref{sec:mapn}.
We compare its speed and numerical agreement with other implementations in Section~\ref{sec:benchmarks}, demonstrate its performance in inference in Section~\ref{sec:performance} and conclude in Section~\ref{sec:conclusion}. 

\section{BiHalofit: Model and \texttt{JAX} Implementation}\label{sec:model}

\subsection{Model}

\begin{figure}[htbp]
\centering
\includegraphics[width=\columnwidth]{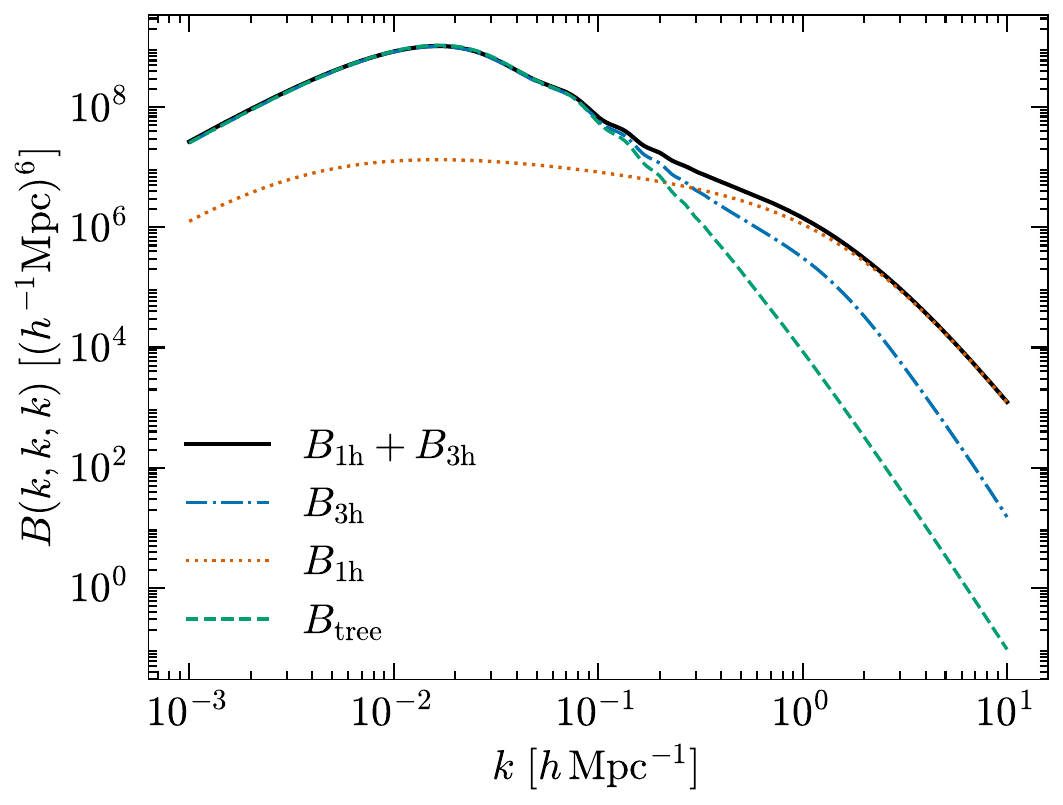}
\caption{The $1$-halo ($B_{1\rm{h}}$) and $3$-halo ($B_{3\rm{h}}$) terms of the BiHalofit fitting formula, for equilateral triangles at $z=0$. The tree level bispectrum ($B_{\rm{tree}}$) is shown for comparison. The cosmological parameters are set to: $\Omega_{\rm cb}=0.3156$, $\Omega_{\rm b}=0.0492$, $h=0.6727$, $n_s=0.9645$, $\sigma_8=0.831$, $M_\nu=0.00\ \rm{eV}$, $w_0=-1$ and $w_a=0$. The linear power spectrum is generated using \texttt{baccoemu}.}
\label{fig:bhf_terms}
\end{figure}

The matter bispectrum is defined as the three-point correlation function of the matter density field in Fourier space:
\begin{equation}
    \langle \delta (\boldsymbol{k}_1)\delta(\boldsymbol{k}_2)\delta(\boldsymbol{k}_3)\rangle = (2\pi)^3 \delta^{(3)}(\boldsymbol{k}_1+\boldsymbol{k}_2+\boldsymbol{k}_3) B_\delta(k_1,k_2,k_3),
\end{equation}
where $\delta(\boldsymbol{k})$ is the Fourier transform of the matter density contrast, $\delta^{(3)}$ is the three-dimensional Dirac delta function (coming from assuming statistical homogeneity of the fields), and $B_\delta(k_1,k_2,k_3)$ is the bispectrum, which depends on the magnitudes of the three wavevectors forming a closed triangle in Fourier space. For large scales, the bispectrum can be computed using perturbation theory, but on smaller scales, nonlinear effects become significant, and one has to resort to fitting formulas or simulations to generate theoretical predictions. BiHalofit \citep{Takahashi2020} provides the current state-of-the-art for modelling these nonlinear scales, providing an analytic fitting formula for the
nonlinear matter bispectrum as a function of cosmology and triangle
configuration. It is parameterised by the nonlinear scale $k_{\rm NL}$ and
effective spectral index $n_{\rm eff}$, defined by
\begin{equation}\label{eq:sigma2_kNL}
  \sigma^2(k_{\rm NL}^{-1}) = 1,
\end{equation}
\begin{equation}\label{eq:sigma2_R}
  \sigma^2(R) = \int \mathrm{d}\ln k\, \Delta^2_{\rm L}(k)\, e^{-k^2 R^2},
\end{equation}
\begin{equation}\label{eq:neff}
  n_{\rm eff} + 3 = -\left.\frac{\mathrm{d}\ln\sigma^2(R)}{\mathrm{d}\ln R}
  \right|_{R\,=\,k_{\rm NL}^{-1}},
\end{equation}
where $\Delta^2_{\rm L}(k) \equiv k^3 P_{\rm L}(k) / (2\pi^2)$ is the dimensionless
linear power spectrum. We suppress the redshift dependence throughout for simplicity. These quantities, together with the amplitude of matter perturbations on scales of $8\,h^{-1}\,{\rm Mpc}$, $\sigma_8$, the energy density fraction of matter,
$\Omega_{\rm m}$, and the tilt of the primordial power spectrum, $n_s$, fully parameterise the fitting formula. The BiHalofit bispectrum combines a $1$-halo and a $3$-halo term,
\begin{equation}
  B_\delta(k_1,k_2,k_3) = B_{\rm 1h}(k_1,k_2,k_3) + B_{\rm 3h}(k_1,k_2,k_3),
\end{equation}
which describe the contribution to the bispectrum from within a halo and from the correlation between three different halos, respectively. Figure~\ref{fig:bhf_terms} shows the two terms separately for equilateral triangles, alongside their sum and the tree-level bispectrum that $B_{\rm 3h}$ reduces to on large scales. The $2$-halo contribution, which is important on intermediate scales and for squeezed triangles ($k_{\rm min}/k_{\rm max}\ll 1$), is neglected in the formula. Writing $q_i \equiv k_i/k_{\rm NL}$ (where $i\in\{1,2,3\}$) the $1$-halo and
$3$-halo terms are parameterised as 
\begin{equation}
  B_{\rm 1h}(k_1,k_2,k_3) = T(q_1)\,T(q_2)\,T(q_3),
\end{equation}
\begin{equation}
  T(q) = \frac{c_n\,q}{\left(a_n q^{\alpha_n} + b_n q^{\beta_n}\right)
  \left(1 + c_n q\right)},
\end{equation}
\begin{equation}\label{eq:B3h}
\begin{split}
  B_{\rm 3h}(k_1,k_2,k_3) &= 2\, I(k_1) I(k_2) I(k_3) 
  \Big[F_2(\boldsymbol{k}_1,\boldsymbol{k}_2) + d_n\,q_3\Big]\\
  &\times P_{\rm E}(k_1)\, P_{\rm E}(k_2) + 2 \ \text{perms.},
\end{split}
\end{equation}
where we have defined
\begin{equation}
  I(k_i) \equiv \frac{1}{1+e_n q_i},
\end{equation}
\begin{equation}\label{eq:Pe}
\begin{split}
  P_{\rm E}(k_i) &\equiv \frac{1 + f_n q_i^2}{1 + g_n q_i + h_n q_i^2}\, P_{\rm L}(k_i) \\
  &+ \frac{(p_n q_i)^3}{m_n q_i^{\mu_n} + n_n q_i^{\nu_n}}
  \frac{1}{1 + (p_n q_i)^3}.
\end{split}
\end{equation}
The permutations in Eq.~(\ref{eq:B3h}) are cyclic over the wavenumbers, and $F_2$ is the second-order perturbation theory kernel,
\begin{equation}
  F_2(\boldsymbol{k}_1,\boldsymbol{k}_2) = \tfrac{5}{7} + \tfrac{1}{2}\cos\theta_{12}
\left(\tfrac{k_1}{k_2}+\tfrac{k_2}{k_1}\right) + \tfrac{2}{7}\cos^2\theta_{12},
\end{equation}
with $\cos\theta_{12}$ fixed by the triangle side lengths via the law of
cosines. The coefficients
($a_n,b_n,c_n,\alpha_n,\beta_n,\gamma_n,d_n,e_n,f_n,g_n,h_n,m_n,n_n,\mu_n,
\nu_n,p_n$) are calibrated against simulations and are fitted power-law functions of $n_{\rm eff}$, $\log\sigma_8$, and the triangle-shape parameters
$r_1 = k_{\rm min}/k_{\rm max}$ and
$r_2 = (k_{\rm mid}+k_{\rm min}-k_{\rm max})/k_{\rm max}$. Appendix~\ref{app:bhf_coeffs} gives their explicit form in terms of the cosmological parameters. We present the calibration range of the fitting function in the first column of Table~\ref{tab:calibration_ranges}. We refer
the reader to \citet{Takahashi2020} for further details about the simulations and fitting procedure used.

\begin{table*}
\centering
\small
\setlength{\tabcolsep}{6pt}
\begin{tabular}{l c c c c c}
\hline\hline
Parameter & BiHalofit & \texttt{baccoemu} & \texttt{syren-new}
 & $\text{BiHalofit}\cap\texttt{baccoemu}$ & $\text{BiHalofit}\cap\texttt{syren-new}$ \\
\hline
$\Omega_{\rm cb}$ & $[0.18,0.45]$ & $[0.06,0.7]$ & $[0.24,0.40]$
 & $[0.18,0.45]$ & $[0.24,0.40]$ \\
$\Omega_{\rm b}$ & $[0.024,0.082]$ & $[0.03,0.07]$ & $[0.04,0.06]$
 & $[0.03,0.07]$ & $[0.04,0.06]$ \\
$h$ & $[0.54,0.93]$ & $[0.5,0.9]$ & $[0.61,0.73]$
 & $[0.54,0.9]$ & $[0.61,0.73]$ \\
$n_s$ & $[0.916,1.013]$ & exact & $[0.92,1.00]$
 & $[0.916,1.013]$ & $[0.92,1.00]$ \\
$\sigma_8$ & $[0.50,1.36]$ & exact & $[0.28,1.30]$
 & $[0.50,1.36]$ & $[0.50,1.30]$ \\
$M_\nu$ [eV] & $0.06$ & $[0,1]$ & $[0,0.15]$ & $0.06$ & $0.06$ \\
$w_0$ & $[-1.2,-0.8]$ & $[-2.0,-0.5]$ & $[-1.3,-0.7]$
 & $[-1.2,-0.8]$ & $[-1.2,-0.8]$ \\
$w_a$ & 0 & $[-0.5,0.5]$ & $[-0.7,0.5]$ & 0 & 0 \\
$z$ & $[0,1.5]$ & $[0,9]$ & $[0,3]$ & $[0,1.5]$ & $[0,1.5]$ \\
$k_{\rm max}$ [$h\,{\rm Mpc}^{-1}$] & $3$ & $50$ & $9$ & $3$ & $3$ \\
\hline
\end{tabular}

\refstepcounter{table}
\label{tab:calibration_ranges}
\begin{minipage}{\textwidth}
\footnotesize
\justifycaption\noindent\textbf{Table \thetable.} Parameter ranges that BiHalofit and the two emulated linear power spectrum backends (Section~\ref{sec:backends}) were fitted or trained on, and, in the last two columns, their intersections: the region where BiHalofit and the backend are both valid. The BiHalofit ranges are the extremes of the sampling box of the 40 $w$CDM simulations of \citet{Takahashi2020}. Note that the simulated cosmologies fill only a thin, curved region of this box, with $\Omega_{\rm m}h^2\in[0.130,0.156]$. The BiHalofit $\sigma_8$ range is converted from $A_s$ with the \texttt{syren-new} formula \citep{Sui_2025}. $\Omega_{\rm cb}=\Omega_{\rm m}-\Omega_\nu$ is the cold matter density; the BiHalofit and \texttt{syren-new} bounds are on $\Omega_{\rm m}$, which differs by $\Omega_\nu\lesssim0.004$. The ``exact'' entries mean the quantity is not bounded. 
\end{minipage}
\end{table*}

\texttt{bihalofit-JAX} also accepts a Chevallier--Polarski--Linder (CPL) dark-energy equation of state, $w(a) = w_0 + w_a(1-a)$. It enters only through the background expansion, which sets $\Omega_{\rm m}(z)$, and the linear growth in the linear power spectrum, so the fitting formula itself is unchanged. The coefficients were calibrated on constant-$w$ simulations, so $w_a \neq 0$ is an extrapolation that has not been validated against simulations; we provide it for generality, e.g.\ for forecasts or priors that vary $w_a$.

In addition to providing a fitting formula for the nonlinear matter bispectrum,
BiHalofit also includes a baryonic-feedback correction, calibrated against
hydrodynamical simulations of IllustrisTNG \citep{2019ComAC...6....2N} and applied
as a per-leg multiplicative boost to the dark-matter-only bispectrum:
\begin{equation}\label{eq:baryon_ratio}
  \frac{B_{\rm b}(k_1,k_2,k_3)}{B_{\delta}(k_1,k_2,k_3)} = R_{\rm b}(k_1)\,R_{\rm b}(k_2)\,
  R_{\rm b}(k_3),
\end{equation}
where $B_{\rm b}$ is the bispectrum including baryonic-feedback. Writing $\xi \equiv \log_{10}k$, the per-leg ratio is
\begin{equation}\label{eq:baryon_leg}
\begin{split}
  R_{\rm b}(k) &= A_0\, e^{-\left(|\xi-\mu_0|/\sigma_0\right)^{\alpha_0}}
  - A_1\, e^{-\left((\xi-\mu_1)/\sigma_1\right)^{2}} \\
  &+ \left[1+(k/k_s)^{\alpha_2}\right]^{\beta_2}.
\end{split}
\end{equation}
The amplitudes $A_0,A_1$, centroids $\mu_0,\mu_1$, widths $\sigma_0,\sigma_1$,
and the shape scale $k_s$ and slope $\beta_2$ are fitted functions of the
scale factor $a$ alone, and the exponents $\alpha_0$ and $\alpha_2$ are constants; Appendix~\ref{app:baryon} gives their explicit form. The
correction is calibrated only up to $z=5$ and is set to $R_{\rm b}=1$ (no
correction) beyond that.
\subsection{Differentiable Linear Power Spectrum Backends}\label{sec:backends}

BiHalofit takes the linear power spectrum $P_{\rm L}(k,z)$ as an input (see Eqs.~(\ref{eq:sigma2_R}) and (\ref{eq:Pe})), traditionally obtained from a Boltzmann-solver such as \texttt{CAMB} \citep{2000ApJ...538..473L,lewis2026cambv2cosmologicalpower} or \texttt{CLASS} \citep{Diego_Blas_2011}. Due to computational cost in inference there is a growing availability of fast, differentiable emulators for $P_{\rm L}$, trained on Boltzmann-solver outputs or simulations, and evaluated accurately orders of magnitude faster \citep{Fendt_2007, fendt2007computinghighaccuracypower,Mootoovaloo_2022, Arico__2022,Spurio_Mancini_2022, Piras_2023,Bartlett_2024,Orjuela-Quintana:2024hha}. In this work we make use of this growing ecosystem by supporting various linear power spectrum backends.

\texttt{bihalofit-JAX} currently supports three 
interchangeable built-in differentiable, fast to evaluate linear power spectrum models. Switching between them requires no other change to the pipeline or user interface. We have designed the backends to be easily extensible, so that users can add their own differentiable linear power spectrum model (e.g.\ a Boltzmann emulator such as \texttt{CosmoPower-JAX} \citep{Piras_2023}) by implementing a single function with the same signature as the built-in backends\footnote{We do not 
currently allow for a user injected table for the linear power spectrum, since one loses differentiability.}. Importantly, the downside of using an emulated linear power spectrum is that the emulator is only valid within its training volume, and will not necessarily extrapolate reliably outside of it. As such we recommend that users check the validity of their cosmology against the training volume of the chosen backend before using it in inference and consult Table~\ref{tab:calibration_ranges} for the training ranges of the two emulated backends.

The three backends, and the trade-offs between them, are:

\begin{description}
\item[1) \texttt{baccoemu}]
  A neural network emulator of the linear cold-matter power spectrum, trained on the outputs of a Boltzmann solver \citep{Arico__2022}, natively supporting massive
  neutrinos and a $w_0w_a$ dark-energy equation of state. Its network evaluates in \texttt{float32}, so we run it in \texttt{float32} throughout the calculation. We call its
  \texttt{JAX}-ported forward pass directly, batching every cosmology
  and redshift into a single call rather than looping over the
  emulator's own Python interface. An important detail is that \texttt{baccoemu} is trained on a fixed $k$-grid of $N_k=600$ logarithmically spaced points over $10^{-4} \le k \le 50 \ h\mathrm{Mpc}^{-1}$, meaning interpolation is required, resulting in a small error which is in principle not present in the other two backends (one can evaluate them at any wavenumber). We discuss this further in Section~\ref{sec:kernel} and Appendix~\ref{app:precision}. Another disadvantage of this emulator is that it uses a Rectified Linear Unit (ReLU) activation function, which leads to discontinuous derivatives in some regions of parameter space. We therefore do not recommend using this backend when computing derivatives, such as in gradient-based inference. 

\item[2) \texttt{syren-new}]
  A symbolic-regression fitting formula for the linear matter power spectrum of \cite{Sui_2025},
  reimplemented here directly in \texttt{JAX} with no external
  dependency\footnote{See \url{https://github.com/DeaglanBartlett/symbolic_pofk/tree/main} for the official implementation.}. Its amplitude is set by converting $\sigma_8$ to $A_s$ with the fitting formula of \citet{Sui_2025}, which is accurate to $\sim0.1\%$, so, as for the other two backends, we then rescale the spectrum to have exactly the requested $\sigma_8$. Its cost within the full calculation is similar to that of \texttt{baccoemu} (Appendix~\ref{app:backends}). Like \texttt{baccoemu}, it natively supports massive
  neutrinos and a $w_0w_a$ dark-energy equation of state. The bounds of its training volume are slightly narrower than \texttt{baccoemu}'s (see Table~\ref{tab:calibration_ranges}), although they match to $\lesssim 1\%$ within the intersecting training region. Note, however, that symbolic-regression emulators are much more successful at extrapolating beyond their training range compared to neural-network emulators. We also highlight that due to the analytical form of the emulator, we have continuous derivatives making this ideal for inference.

\item[3) Eisenstein \& Hu (1998)]
  The analytic Eisenstein \& Hu (1998) no-wiggle transfer function (E\&H from hereon)
  \citep{EisensteinHu1998}. It ignores massive neutrinos (though it
  still accepts a $w_0,w_a$ equation of state) and is the least
  accurate of the three, but it is dependency-free
  and useful as a minimal fallback or for
  isolating BiHalofit-specific behaviour from emulator or
  symbolic-regression systematics.
\end{description}

We quantify the effect of the backend choice on the final statistics, holding cosmology, the BiHalofit fitting formula and the triangle set fixed, in Appendix~\ref{app:backends}. Relative to \texttt{syren-new}, E\&H deviates in the matter bispectrum by a mean of $3\%$ (at most $18\%$), while \texttt{baccoemu} deviates by a mean of $0.2\%$ (at most $3.3\%$), over $1000$ cosmologies in the region the simulations cover. The differences between \texttt{syren-new} and \texttt{baccoemu} are small compared with the accuracy of the BiHalofit fitting formula itself of up to $\sim 20\%$ \citep{Takahashi2020}, whereas E\&H is the least accurate of the three.

\subsection{\texttt{JAX} kernel design}\label{sec:kernel}
Where the original implementation\footnote{\url{http://cosmo.phys.hirosaki-u.ac.jp/takahasi/codes_e.htm}} of BiHalofit  \citep{Takahashi2020} locates $k_{\rm
NL}$ by bisection against an adaptively integrated $\sigma^2(R)$,
independently per redshift and cosmology, our kernel instead uses a
safeguarded (bisection-backed) Newton iteration on
Eq.~(\ref{eq:sigma2_kNL}), evaluated directly on a fixed $k$-grid and
batched across cosmologies and redshifts. The same Newton derivative then evaluates the $n_{\rm eff}$ integral of Eq.~(\ref{eq:neff}). This is a more efficient root-finding method, which converges in around $N_R\sim10$ steps (compared to $\sim 17$ steps for the bisection-based method) for both \texttt{float32} and \texttt{float64}\footnote{When computing with a floating-point precision we maintain that precision throughout the calculation to avoid mixed-precision issues in different versions of \texttt{JAX}.}. 
In addition to fewer iterations, our method is directly optimised for GPU work, where the number of iterations is fixed to avoid host-to-device transfers in operations such as while loops which slow down processes. The kernel also returns the size of the last Newton step, which the code can check against a user-supplied tolerance to validate the convergence. The convergence tests and recommended precision settings are examined in Appendix~\ref{app:precision}.

The linear power spectrum is tabulated on a logarithmically spaced $k$ grid to evaluate the integrals of Eqs.~(\ref{eq:sigma2_R}) and (\ref{eq:neff}). This is done independently for each redshift slice. To avoid further evaluations of the linear power spectrum, we perform linear interpolation of this table in $\log P_\mathrm{L}$ against $\log k$, and use the interpolant to evaluate Eq.~(\ref{eq:Pe}). A resolution of $N_k=600$ for $10^{-4} \le k \le 50\,h\,\mathrm{Mpc}^{-1}$ keeps the interpolation error on the bispectrum 
at $0.019\%$ on average and at most $0.18\%$, below the mean difference of $0.2\%$ between the more accurate linear power spectrum backends, \texttt{syren-new} and \texttt{baccoemu} (Appendix~\ref{app:backends}).
These performance parameters can be tuned in our code to trade off speed and accuracy, which we examine in more detail in Appendix~\ref{app:precision}. We show in Figure~\ref{fig:make_bihalofit_example} the full calling signature,
building the model once for a fixed triangle/redshift grid and evaluating the bispectrum at one cosmology.
\begin{figure}[htbp]
\begin{lstlisting}[style=pythonstyle]
# Import packages
import numpy as np
from bihalofit_jax import make_bihalofit, Cosmology

# Build cosmology
cosmo = Cosmology(
    omega_cold=np.array([0.3156]),
    omega_baryon=np.array([0.0492]),
    sigma8_cold=np.array([0.831]),
    ns=np.array([0.9645]),
    hubble=np.array([0.6727]),
    neutrino_mass=np.array([0.0]),  
    w0=np.array([-1.0]),
    wa=np.array([0.0]),
)

# Build JIT-compiled bihalofit kernel
k = np.array([0.1, 0.2, 0.3])
bhf = make_bihalofit(
    k1=k, k2=k, k3=k,
    expfactors=np.array([1.0]),  
    pk_source='baccoemu',  
    N_R=15,                
    tol=None,               
    R_min=1e-6,
    R_max=40.0,             
    dtype=np.float32,
    validate=False,         
    term='full',            
    baryonic_boost=False,   
    k_lin=None,             
)

# Evaluate bispectrum at the cosmology
B = bhf(cosmo)
\end{lstlisting}
\caption{Example of generating a matter bispectrum prediction using \texttt{bihalofit-JAX}. Tutorials are provided alongside the package, including a full example of the weak-lensing bispectrum, skew spectra and mass-aperture statistics.}
\label{fig:make_bihalofit_example}
\end{figure}

The full kernel is compiled with a
single just-in-time (JIT) compilation: cosmologies are fully batched, the redshift axis makes use of the \texttt{vmap} tool  of \texttt{JAX}, and triangle configurations are handled by broadcasting. This design allows the bispectrum to be computed in $\sim 1$--$2\,\text{ms}$, far faster than any other implementation of BiHalofit, as we show in Section~\ref{sec:timing_matter}. Not only this, but it also scales extremely well with the number of cosmologies and redshifts, making it suitable for inference pipelines that require the evaluation of the bispectrum at many cosmologies and redshifts, or for batch evaluations of the formula for training emulators. The kernel can be evaluated in \texttt{float32} or \texttt{float64}, with the former being the default. This is critical for GPU performance, since GPUs (especially modern ones) are optimised for \texttt{float32} operations. We find that the loss of precision in \texttt{float32} does not significantly affect the accuracy of the bispectrum, and the speed-up gained from using \texttt{float32} is significant. The design of this base kernel enables the efficient computation of downstream statistics, such as the weak-lensing bispectrum, skew spectra and mass-aperture statistics, since one is required to evaluate the bispectrum many times across redshift and wavenumbers, as we show in Sections~\ref{sec:projection} and \ref{sec:performance}. 

\section{Projected Summary Statistics}\label{sec:projection}

\subsection{Projected Bispectra}\label{sec:bkappa}
In photometric surveys we observe different tracers of the large scale structure of the Universe 
which are projected along the line of sight. These tracers might include for example galaxy clustering, weak lensing, or CMB lensing. 
Statistical analyses including correlations of these observables (e.g. angular power spectra or bispectra) have been shown to be a powerful probe of cosmology \citep{Garcia-Garcia:2021unp,descollaboration2026darkenergysurveyyear}. In this section we show how to compute these projected statistics using the BiHalofit kernel.
It can be shown that the projected bispectrum (examples include
galaxy-galaxy-lensing or lensing-lensing-lensing bispectra) is related to the three-dimensional matter bispectrum in the Limber approximation \citep{1953ApJ...117..134L,LoVerde_2008} (valid for large $\ell\gg 1$) 
\begin{equation}\label{eq:projected_bispectra}
\begin{split}
  B_{ijk}(\ell_1,\ell_2,\ell_3) &= \int \frac{\mathrm{d}\chi}{\chi^4}
  W_i(\chi)W_j(\chi)W_k(\chi)\times\\
  & \times B_\delta\left(k_{\ell_1}, k_{\ell_2}, k_{\ell_3}; z(\chi)\right),
\end{split}
\end{equation}
where $k_{\ell_i}\equiv (\ell_i+1/2)/\chi$ and $W_i(\chi)$ is the kernel corresponding to a given tracer. The matter bispectrum can be computed on nonlinear scales using the BiHalofit model\footnote{At small $\chi$, Eq.~(\ref{eq:projected_bispectra}) samples scales far smaller than the $k_{\rm max}=3h{\rm Mpc}^{-1}$ to which BiHalofit was calibrated (Table~\ref{tab:calibration_ranges}). These contributions are suppressed by the kernels but use the fitting formula outside its range of validity.}, which is described in Section~\ref{sec:model}. In the case of galaxy-clustering the kernel is
\begin{equation}\label{eq:wcounts}
  W_{\rm counts}(\chi) = b\,n_g(z(\chi))\,H(z(\chi))
\end{equation}
for a linear bias, $b$. In weak lensing (cosmic shear), one considers the coherent distortion of light rays from source galaxies caused by the large scale structure of the Universe. With a source distribution $n(z)$, the kernel is described by 
\begin{equation}\label{eq:wlensing}
  W_{\rm lensing}(\chi) = \frac{3}{2}\,\Omega_{\rm m}H_0^2
  \frac{\chi}{a(\chi)}\int_{z(\chi)}^{\infty}\mathrm{d}z' n(z')\,\frac{\chi(z') - \chi}{\chi(z')}\,,
\end{equation}
where we have set $c=1$. Note that in the full sky there is usually a factor of $\sqrt{(\ell+2)!/(\ell-2)!}(\ell+1/2)^{-2}$ which appears in the lensing kernel; however we neglect this in the first version of our code since it is negligibly close to unity for $\ell\geq 10$ (where the Limber approximation is valid). In so-called beyond-Limber statistics these factors would become important on large angular scales. The CMB-lensing kernel has the same functional form but where $n(z)$ is now a delta function at
$z_{\rm source}\approx1089$. We write $B_\gamma$ for the projected weak-lensing bispectrum, i.e.\ Eq.~(\ref{eq:projected_bispectra}) with the lensing kernel on all three legs.

A very important effect for cosmic shear, particularly in stage IV surveys, is the impact of intrinsic alignments (IA) of the source galaxies with their local environment \citep{Brown_2002}. This is typically taken into account using the nonlinear alignment (NLA) model \citep{HirataSeljak2004,BridleKing2007}, which results in the kernel
\begin{align}\label{eq:wia}
  W_{\rm IA}(\chi) &= f_{\rm IA}(z(\chi))\,n(z(\chi))\,H(z(\chi))\,, \\
  f_{\rm IA}(z) &= -A_{\rm IA}\,C_1\rho_{\rm crit}\,\frac{\Omega_{\rm m}}{D(z)}
  \left(\frac{1+z}{1+z_{\rm pivot}}\right)^{\eta}\,, \nonumber
\end{align}
where $n(z)$ is the source distribution, $D(z)$ the linear growth factor normalised to $D(0)=1$, $C_1\rho_{\rm crit}=0.0134$, $z_{\rm pivot}=0.62$ and $A_{\rm IA}$ and $\eta$ are free parameters.

We evaluate the Limber integral, Eq.~(\ref{eq:projected_bispectra}), with Gauss--Legendre (GL) quadrature in $\chi$. By default we use $N_\chi=160$ nodes over $10\le\chi\le5800\,h^{-1}{\rm Mpc}$. The nodes are fixed and shared by all three kernels, so the integral needs one BiHalofit evaluation per node. This is directly compatible with the BiHalofit kernel design, which is optimised for many evaluations of the bispectrum at once. 

Each GL node requires the redshift $z(\chi_j)$ to evaluate the matter bispectrum (BiHalofit), so we calculate (as a default) $\chi(z)$ on a grid of $N_z^{\rm{bg}}=2000$ uniformly spaced redshifts up to $z=10$, and invert it at the nodes using a Catmull--Rom spline (this keeps the result differentiable as the mapping from comoving distance to redshift depends on the cosmology). The clustering and IA kernels also need the evaluation $n(z(\chi))$ at the same comoving distances, so we interpolate (again using a Catmull--Rom spline) the user-inputted $n(z)$ to the node redshifts $z(\chi_j)$. With the default settings the mean error is $\lesssim0.001\%$ for lensing and IA kernels and $\lesssim0.05\%$ with clustering legs (Appendix~\ref{app:bkappa}).

\subsection{Skew Spectra}\label{sec:skew}

Skew spectra compress the bispectrum's full triangle dependence into a
power-spectrum-like function of a single multipole, by cross-correlating a
field with a composite (squared/quadratic) map built from another field and
fixing one leg of the underlying bispectrum rather than binning all three.
A simple example is to construct the square of the (filtered) galaxy overdensity field, $g$, and take the correlation of this with the original field, from which one can show that 
\begin{equation}\label{eq:skew_spectrum}
  C_{\ell}^{g^2 g} \equiv \sum_{\ell_1,\ell_2} \frac{(2\ell_1+1)(2\ell_2+1)}{4\pi} \begin{pmatrix}
  \ell_1 & \ell_2 & \ell\\
  0 & 0 & 0
  \end{pmatrix}^2 B_{ggg}(\ell,\ell_1,\ell_2), 
\end{equation}
where $\langle (g^{2})_{\ell m}g_{\ell'm'}^{*}\rangle \equiv \delta^{\mathcal K}_{\ell \ell'}\delta^{\mathcal K}_{mm'}C_{\ell}^{g^2 g}$ defines the angular power spectrum between the two fields (the skew spectrum) and the matrix is the Wigner-$3j$ symbol which encodes the angular momentum conservation on the sphere\footnote{The galaxy overdensity is first filtered before squaring to ensure that the sum is over a finite number of multipoles.}. Here we have that $B_{ggg}$ refers to the angular bispectrum computed using Eq.~(\ref{eq:projected_bispectra}) with the number counts kernel Eq.~(\ref{eq:wcounts}) on each leg (we suppress the redshift information for simplicity). More general constructions are possible, including cross-correlations between different fields and different quadratic combinations of the fields, including filtered combinations of fields \citep{Harscouet2025Fast, Harscouet2026Constraints,Verdiani:2025jgz,Roskill:2025eot}. These statistics have the key advantage of leveraging 
the highly developed infrastructure surrounding the efficient estimation of power spectra \citep{Alonso_2019} and their covariances \citep{Harscouet2025Fast}, while retaining much of the information content of the bispectrum. Despite this, the modelling of these statistics is still 
a significant challenge, particularly with regards to computational expense, that we aim to ameliorate here.

To keep our implementation of skew spectra relatively general we implement the formula 
\begin{equation}
  \begin{split}
    C_{\ell}^{(jk)i} &= \sum_{\ell_1,\ell_2} \frac{(2\ell_1+1)(2\ell_2+1)}{4\pi}  M_{a}(\ell,\ell_1,\ell_2)M_{b}(\ell,\ell_1,\ell_2) \\
    &\times F_{\ell_1}F_{\ell_2}B_{ijk}(\ell,\ell_1,\ell_2),
  \end{split}
  \label{eq:skew_general}
\end{equation}
where $M_a$ (and equivalently $M_b$) is the Wigner-$3j$ symbol
\begin{equation}
  M_{a} = \begin{pmatrix}
    \ell_1 & \ell_2 & \ell\\
    s_{1,a} & s_{2,a} & s_{a}
    \end{pmatrix},
\end{equation}
where $s_{1,a}$, $s_{2,a}$ and $s_a$ specify the spin-weights\footnote{A spin-weighted field, $f$, of spin $s$,  is defined through its transformation under a rotation of $\psi$: $f\rightarrow e^{is\psi}f$. See \citet{Roskill:2025eot} for further details.} of the fields being considered, satisfying $s_{1,a}+ s_{2,a} +s_{a}=0$ (conservation of spin). This lets the user compute spin-weighted versions of the skew spectra statistics, which were introduced recently in \citet{Roskill:2025eot}.
The user can provide a given bispectrum of Eq.~(\ref{eq:projected_bispectra}) and can choose which filters to include on the two legs of the bispectrum, encoded in $F_{\ell_1}$ and $F_{\ell_2}$. The filters can be used to isolate different scales of interest. We precompute the Wigner-$3j$ symbols needed for the skew spectra using \texttt{pyshtools} \citep{https://doi.org/10.1029/2018GC007529}; this is a fixed cost independent of the main calculation. 
The filters are most commonly top-hats over a band
$[\ell_{\rm lo},\ell_{\rm hi}]$,
\begin{equation}
  F^{\rm TH}_{\ell} =
  \begin{cases}
    1, & \ell_{\rm lo} \le \ell \le \ell_{\rm hi}\\
    0, & \mathrm{otherwise},
  \end{cases}
  \label{eq:tophat}
\end{equation}
selecting the scales retained in the squared field. Since one might be interested in differentiating with respect to the filters, e.g. to optimise the Fisher information of these statistics, we also provide a smooth band, 
\begin{equation}
  F_{\ell}(\ell_{\rm lo},\ell_{\rm hi},w) =
    \sigma\!\left(\frac{\ell-\ell_{\rm lo}+\tfrac{1}{2}}{w}\right)
    \sigma\!\left(\frac{\ell_{\rm hi}+\tfrac{1}{2}-\ell}{w}\right),
  \label{eq:soft_tophat}
\end{equation}
with $\sigma(x)=(1+e^{-x})^{-1}$ and $w$ a ramp width in multipoles. We make it straightforward to implement other filters. 

Evaluating the skew spectra requires a double sum over $\ell_1$ and $\ell_2$, which is particularly expensive for wide filters. To speed it up, we provide the option to interpolate the bispectrum on a grid of $\ell_1$ and $\ell_2$ values chosen by the user, since the best choice depends on the scales the filter probes and on the accuracy required.  Unlike for the other statistics, we do not recommend precision settings, instead we provide a detailed tutorial showing how to choose them for specific cases.

\subsection{Aperture Statistics}\label{sec:mapn}

\texttt{bihalofit-JAX} also provides support for third-order aperture
statistics \citep{Schneider1998, Crittenden2002, Linke:2024huv}. Examples are the mass-aperture statistics, which are built from the weak-lensing convergence of a tomographic bin $i$ at sky position $\boldsymbol{\varphi}$, the line-of-sight integral of the matter overdensity weighted by the lensing kernel of Eq.~(\ref{eq:wlensing}) (in the flat-sky approximation),
\begin{equation}
  \kappa^i(\boldsymbol{\varphi}) = \int \mathrm{d}\chi\, W^i_{\rm lensing}(\chi)\, \delta(\chi\boldsymbol{\varphi},\chi),
\end{equation}
where $\delta(\boldsymbol{x},\chi)$ is the matter density contrast at transverse comoving position $\boldsymbol{x}$ and comoving distance $\chi$. The third-order aperture-mass statistic is the three-point correlation function of the aperture mass, $M_{\rm ap}$, a filtered version of the convergence field for a given aperture scale $\theta$:
\begin{equation}
M_{\rm{ap}}^{i}(\boldsymbol{\varphi},\theta)=\int \mathrm{d}^2 \boldsymbol{\varphi}' \, u_{\theta}(|\boldsymbol{\varphi}-\boldsymbol{\varphi}'|,\theta) \kappa^i(\boldsymbol{\varphi}'),
\end{equation}
where $u_{\theta}$ is the filter. One can show that 
\begin{equation}\label{eq:mass_aperture}
  \begin{split}
  \langle M_{\rm ap}^3\rangle(\theta_1, \theta_2, \theta_3) &= \int \frac{\mathrm{d}^2\ell_1}{(2\pi)^2} \frac{\mathrm{d}^2\ell_2}{(2\pi)^2} B_\gamma(\ell_1,\ell_2,|\boldsymbol{\ell_1+\ell_2}|)\\
  &\times \hat u(\ell_1\theta_1)\hat u(\ell_2\theta_2)\hat u(|\boldsymbol{\ell_1+\ell_2}|\theta_3),
  \end{split}
\end{equation}
where $\hat u$ is the Fourier-space representation of the filter and $B_{\gamma}$ is the weak lensing convergence bispectrum (in the flat sky approximation) defined in Eq.~(\ref{eq:projected_bispectra}). We provide functionality for the
\citet{Crittenden2002} aperture filter\footnote{We plan to implement other filter types in future versions.}, given by $\hat u(\eta) = \tfrac{1}{2}\eta^2 e^{-\eta^2/2}$. We implement a general third-order aperture statistic (see e.g. \citet{Schneider_2005,2023ascl.soft06054H}) by allowing for mixed tracer objects
\begin{align}
  \langle X_1 X_2 X_3 \rangle(\theta_1, \theta_2, \theta_3) =
  \frac{2}{8\pi^3} \int_0^\infty \mathrm{d}\ell_1 \int_0^\infty
  \mathrm{d}\ell_2 \int_0^\pi \mathrm{d}\varphi\; \ell_1 \ell_2\, & \nonumber\\
  B(\ell_1,\ell_2,\ell_3)\,
  \hat u(\ell_1\theta_1)\, \hat u(\ell_2\theta_2)\, \hat u(\ell_3\theta_3)&,
\end{align}
where $\ell_3=\sqrt{\ell_1^2+\ell_2^2+2\ell_1\ell_2\cos\varphi}$ closes the
Fourier triangle and $\varphi=\phi_2-\phi_1$ is the relative angle. Note that we have re-written the integration variables to be in terms of $\varphi$.

\begin{figure*}[htbp]
\centering
\includegraphics[width=\textwidth]{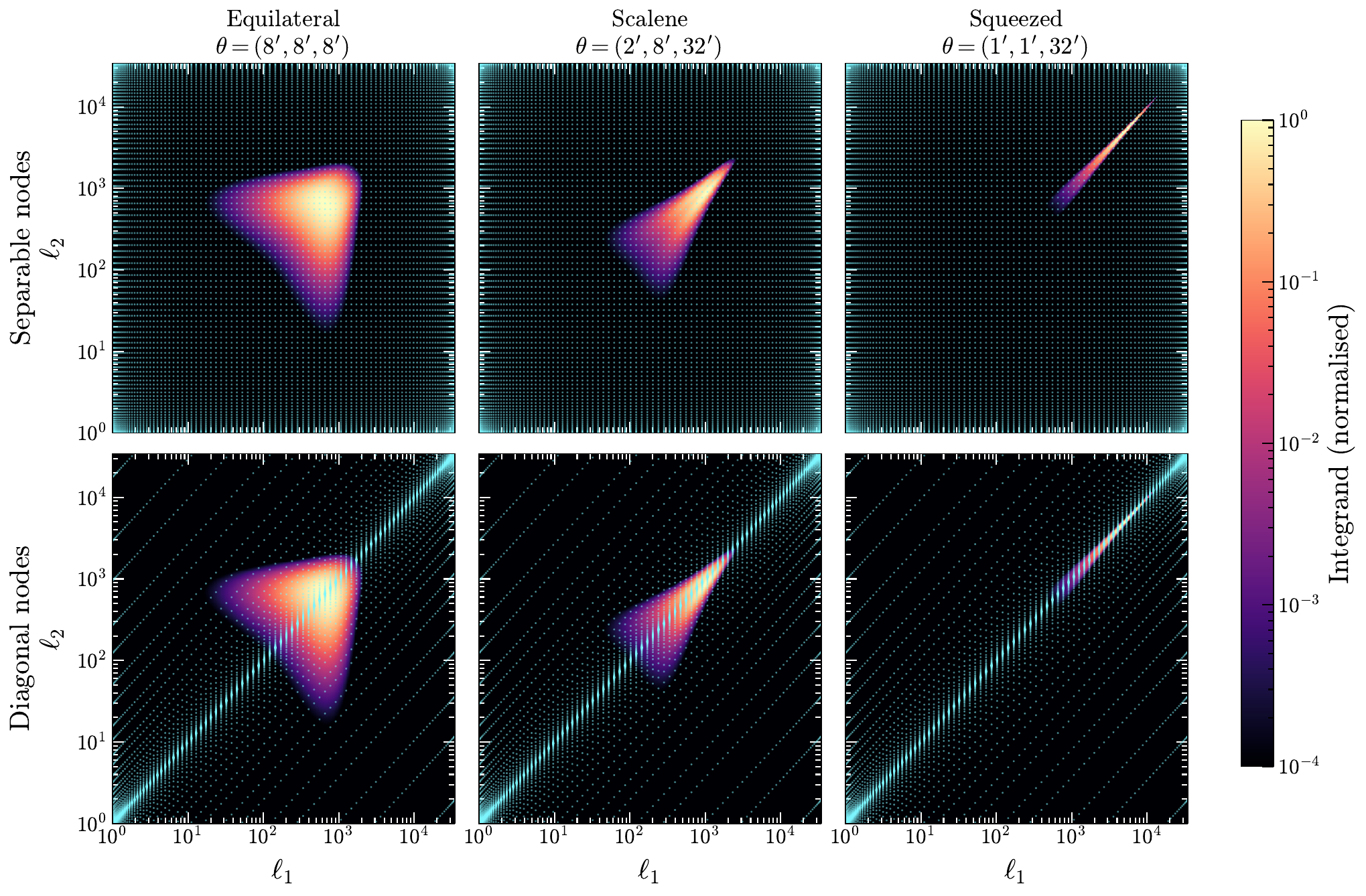}
\caption{The $\langle M_{\rm ap}^3\rangle$ integrand in the $(\ell_1,\ell_2)$
plane, with the bispectrum set to unity and $\varphi$ integrated out on a fine uniform grid, for an
equilateral, a scalene and a squeezed aperture triplet (left to right).
Quadrature nodes are overlaid in cyan: sampled on $\ell_1$ and $\ell_2$
independently (top) and on $\ell_1$ and the ratio $\rho=\ell_2/\ell_1$
(bottom).}
\label{fig:map3_grid_choice}
\end{figure*}

In the case of the Crittenden filter, the integrand
is a product of three sharply peaked functions: $\hat u(\eta)$ is maximal at
$\eta = \sqrt{2}$, so leg $i$ draws its signal from
$\ell_i \simeq \sqrt{2}/\theta_i$. Only two of those peaks can be targeted directly in its numerical computation. Once $\ell_1$ and
$\ell_2$ are chosen, $\ell_3$ follows from the triangle closure, and for a `squeezed' configuration such as
$\theta_3 \gg \theta_1, \theta_2$ the third filter needs
$\ell_3 \simeq \sqrt{2}/\theta_3 \ll \ell_1$, which the inequality
$\ell_3 \geq |\ell_1 - \ell_2|$ permits only for
\begin{equation}
  \left|\ell_1 -\ell_2\right| \lesssim \frac{\sqrt{2}}{\theta_3}.
  \label{eq:map3_band}
\end{equation}
Given that $\theta_3$ is large, the signal therefore resides in a region around
$\ell_2 = \ell_1$, which narrows as $\ell_1$ and $\ell_2$ increase. This makes a fixed grid on $\ell_1$ and $\ell_2$ highly inaccurate for squeezed configurations, prompting codes to use an adaptive integration scheme that refines the grid where more resolution is needed. This is shown for a given grid of $\ell_1$ and $\ell_2$ in the top row of Figure~\ref{fig:map3_grid_choice}. 

We would like to work with a fixed grid, rather than use an adaptive scheme since we want to limit as far as possible the host-to-device transfers which slow down our code and make it less useful for GPU work. To make this possible we replace the $\ell_2$ integration variable by the ratio
${\rho \equiv \ell_2/\ell_1}$, such that the integral becomes
\begin{align}
  \langle X_1 X_2 X_3 \rangle =
  \frac{2}{8\pi^3} \int_0^\infty\! \mathrm{d}\ell_1 \int_0^\infty\!
  \mathrm{d}\rho \int_0^\pi\! \mathrm{d}\varphi\; \ell_1^3\, \rho\,
  & B(\ell_1, \rho\ell_1, \ell_3) \nonumber\\
  \times\, \hat u(\ell_1\theta_1)\, \hat u(\rho\ell_1\theta_2)\,
  \hat u(\ell_3\theta_3)&,
  \label{eq:map3_diagonal}
\end{align}
with $\ell_3 = \ell_1\sqrt{1+\rho^2+2\rho\cos\varphi}$.
Figure~\ref{fig:map3_grid_choice} shows why this is the natural coordinate. The thin region of signal now sits
at the fixed, known location $\rho = 1$ instead of moving with $\ell_1$ and
$\theta_3$, and is covered by construction: $\ell_1$ carries a global
GL rule of $N_\ell$ nodes in $\ln\ell_1$ over $1\le\ell_1\le10/\theta_{\min}$, with $\theta_{\min}$ the smallest aperture, while the $\rho$ nodes are
placed in $N_\rho$ panels of two GL nodes each, spaced geometrically in $|\ln\rho|$ from $3\times10^{-3}$ to $\ln(10/\theta_{\min})$ and mirrored about
$\rho = 1$, which
spans Eq.~(\ref{eq:map3_band}) over the whole $\ell_1$ range, and $\varphi$ carries a GL rule of $N_\varphi$ nodes over $[0,\pi]$. By default we set $N_\ell=96$, $N_\rho=48$ and $N_\varphi=64$ (Appendix~\ref{app:map3}). This setup enables a straightforward JIT compilation and fast computation in \texttt{JAX}, forming the key numerical improvement compared to other implementations which we show explicitly in Section~\ref{sec:benchmarks}.

\section{Benchmarks}\label{sec:benchmarks}

\subsection{Benchmark setup}\label{sec:implementations}

\subsubsection{Comparison codes and settings}\label{sec:comparison_codes}
We compare \texttt{bihalofit-JAX} with independent implementations of the BiHalofit fitting formula\footnote{We have also compared with other implementations privately; here we restrict to those used in published work.} in speed and numerical precision. This is distinct from the physical accuracy, which is set by the formula itself. We include in this comparison: the original code of \citet{Takahashi2020} (referred to hereafter as T20); \texttt{fastnc}\footnote{\url{https://github.com/git-sunao/fastnc}} \citep{sugiyama2024fastmodelingshearthreepoint}; the \texttt{C++} (CPU) and CUDA (GPU) builds of \texttt{threepoint} \citep{2023ascl.soft06054H,Heydenreich2023}\footnote{\url{https://github.com/sheydenreich/threepoint}}; and its Python/\texttt{numba} port \texttt{threepoint\_py}\footnote{\url{https://github.com/llinke1/threepoint_py}} \citep{2026zndo..19072610L}. They differ in how they locate $k_{\rm NL}$ and $n_{\rm eff}$ (Eqs.~\ref{eq:sigma2_kNL}--\ref{eq:neff}) and in how they batch and parallelise the calculation. T20 is a standalone program whose cosmology and redshift are fixed at compile time, so we compile it as a library with added entry points that set the cosmology and pass the linear power spectrum as a table; the fitting formula is unchanged. The other codes are driven through their public interfaces.

We use an NVIDIA RTX PRO 6000 Blackwell GPU and an AMD EPYC 9654 CPU, restricted to $24$ of its $96$ cores, with \texttt{JAX} 0.4.38, CUDA 12.9 for the GPU build of \texttt{threepoint} and GCC 11.5 for the C and C\texttt{++} codes. 

Only \texttt{bihalofit-JAX} evaluates several cosmologies in one call. The other codes run one cosmology at a time, so their time for $N_{\rm cosmo}$ cosmologies is $N_{\rm cosmo}$ times that for one\footnote{They could also run several cosmologies in parallel, for instance with MPI or as separate processes; for the codes that use one core this could divide their time for many cosmologies by up to the $24$ cores of our CPU runs. We leave this out of the comparison for simplicity.}. Within a cosmology, \texttt{threepoint} uses OpenMP in its set-up and aperture statistics. \texttt{fastnc} uses threads for $\langle M_{\rm ap}^3\rangle$ and one core for the bispectra, and \texttt{threepoint\_py} and T20 run on one core.

We use \texttt{syren-new} as the linear power spectrum throughout (except for \texttt{threepoint\_py} which computes its own E\&H spectrum), since it runs in both \texttt{float32} and \texttt{float64} and so allows comparisons at the precision of the codes that run only in \texttt{float64}. The backends are compared in Appendix~\ref{app:backends}. We exclude build and compilation times, since this is a one-off cost that is negligible over an MCMC run: for \texttt{bihalofit-JAX} it is $1$--$3\,{\rm s}$ for the matter bispectrum, $4$--$7\,{\rm s}$ for $B_\gamma$ and $25$--$40\,{\rm s}$ for the skew spectra and $\langle M_{\rm ap}^3\rangle$. We include per-cosmology set-up, which an MCMC run pays at every step, such as the bispectrum precomputation of \texttt{threepoint\_py} and the start-up of \texttt{threepoint}, which runs one process per cosmology. The timing of \texttt{bihalofit-JAX} includes evaluating the linear power spectrum. The codes that accept one (T20, \texttt{fastnc} and \texttt{threepoint}) are instead given each cosmology's $z=0$ \texttt{syren-new} spectrum as a precomputed table on the default $k$ grid of \texttt{bihalofit-JAX} (below), which is not included in their cost. \texttt{threepoint\_py} cannot take a table and computes its own E\&H spectrum as part of its set-up.

With regard to precision, \texttt{bihalofit-JAX} is timed at the following settings: $N_R=15$ Newton iterations, $N_\chi=160$ GL nodes over $10\le\chi\le5800\,h^{-1}{\rm Mpc}$, $N_z^{\rm bg}=2000$ uniformly spaced redshifts up to $z=10$ for $\chi(z)$, and $N_k=600$ log-spaced points over $10^{-4}\le k\le50\,h\,{\rm Mpc}^{-1}$. Every code reads $P_{\rm L}$ on this grid, in the timings and in the accuracy comparisons. The other codes use their default settings. Some of these are less accurate than typical settings of \texttt{bihalofit-JAX} (Appendix~\ref{app:precision}), and matching its accuracy would make them slower. We keep these defaults because a lower numerical precision can still be subdominant to the $\sim 20\%$ physical accuracy of BiHalofit (making it a reasonable choice), and to avoid modifying core functions of the other codes. The other codes evaluate the Limber approximation at $k=\ell/\chi$ (flat sky), which we use in \texttt{bihalofit-JAX} for the sake of comparison, although it is possible to use $k=(\ell+1/2)/\chi$ (curved sky) instead.

\subsubsection{Cosmologies and source distributions}\label{sec:cosmologies}
All timings and comparisons use one set of $1000$ cosmologies, drawn as a Latin hypercube over the sampling box of the simulations on which BiHalofit was calibrated \citep{Takahashi2020} and kept only where they lie inside the training range of \texttt{syren-new} (last column of Table~\ref{tab:calibration_ranges}, with massless neutrinos). This is with the exception of the CPU version of \texttt{threepoint}, whose projection assumes a $\Lambda$CDM expansion history: it is compared at $w_0=-1$. A timing with $N_{\rm cosmo}$ cosmologies uses the first $N_{\rm cosmo}$, and the codes that run one cosmology at a time are timed on the first ($\Omega_{\rm cb}=0.298$, $\Omega_{\rm b}=0.047$, $h=0.696$, $\sigma_8=0.751$, $n_s=0.937$, $w_0=-1.03$). The matter bispectrum is compared on all $1000$ cosmologies and the projected statistics on the first $30$ (calculations are more expensive so we avoid making as many calls), each with its own Smail source distribution,
\begin{equation}\label{eq:Smail}
  n(z)\propto(z/z_0)^\alpha\exp[-(z/z_0)^\beta],
\end{equation}
with $(z_0,\alpha,\beta)$ drawn over $z_0 \in [0.15,0.60]$, $\alpha \in [1,3]$ and $\beta\in [0.8,2]$. The same $n(z)$ is used on all three legs, tabulated from $z=0$ to $z=5$ on $1001$ linearly spaced points and passed to every code. Timings of the lensing statistics use the first cosmology's $n(z)$ for all $N_{\rm cosmo}$ cosmologies, since \texttt{bihalofit-JAX} fixes the source distribution at compilation. The results in this section are therefore valid for this set of source distributions, and care must be taken particularly when using very narrow or high redshift distributions. 

\subsection{Matter bispectrum}\label{sec:timing_matter}\label{sec:agreement_matter}

We time the matter bispectrum at $z=1$ for a fixed set of $10{,}160$ triangles\footnote{The cost is approximately independent of the triangle shape.} (triangles formed from $75$ logarithmically-spaced wavenumbers between $0.01$ and $3\,h\,{\rm Mpc}^{-1}$), for one and for $500$ cosmologies (described in Section~\ref{sec:cosmologies}). For one cosmology \texttt{bihalofit-JAX} takes $\sim 0.932 \ \text{ms}$ and $\sim 1.4\,\text{ms}$ on CPU and GPU respectively in \texttt{float32}, $\sim 15$ times faster than the fastest other implementation, T20, and $\sim 10^3$ times faster than the slowest (\texttt{threepoint\_py}). At this scale the GPU is slower than the CPU because the overhead of launching the kernel dominates the cost of the calculation. The advantage of having a GPU compatible code becomes clear when evaluating many cosmologies at once, where the GPU can batch the calculations: for $500$ cosmologies it takes $\sim 2\,\text{ms}$ on GPU ($\sim 60\,\text{ms}$ on CPU), against $\sim 7\,\text{s}$ for T20, a speed-up of $\sim 4\times10^3$, and $\sim 5\times10^5$ against \texttt{threepoint\_py}. These results are summarised in Table~\ref{tab:projected_speeds}.

\begin{figure*}[htbp]
\centering
\includegraphics[width=\textwidth]{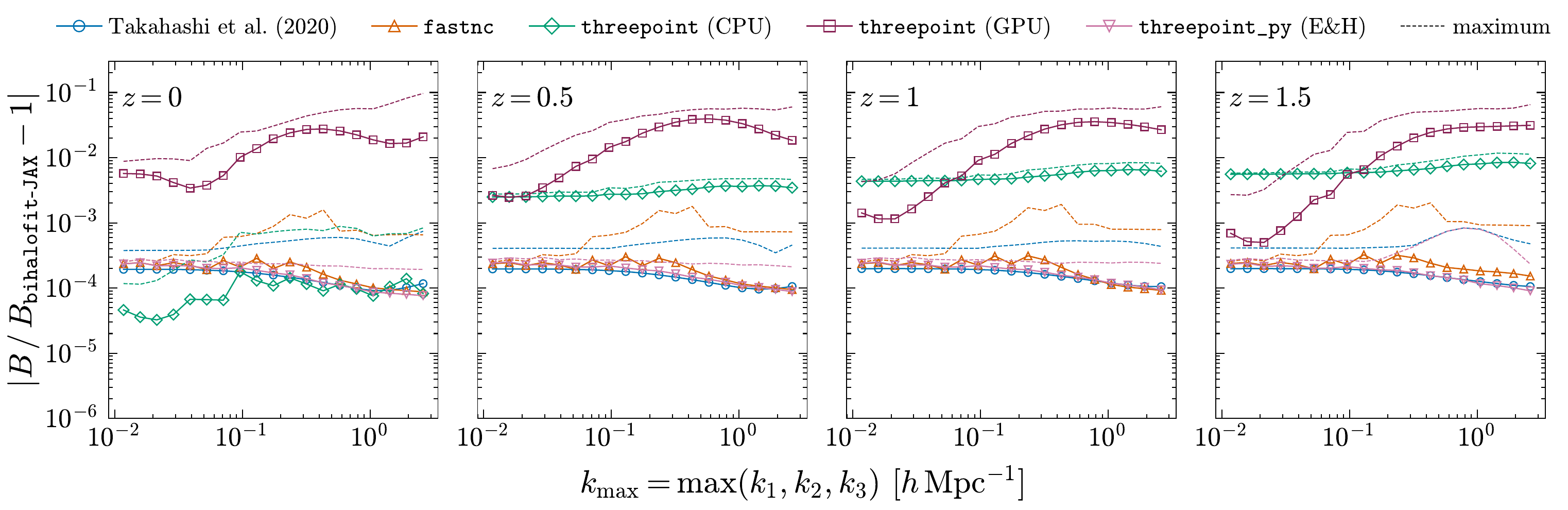}
\caption{Mean (solid) and maximum (dashed) fractional differences between independent implementations and \texttt{bihalofit-JAX} against $k_{\rm max}=\max(k_1,k_2,k_3)$. This is shown for an average over the $1000$ cosmologies of the benchmark set and the triangles in each $k_{\rm max}$ bin, at four redshifts ($z\in\{0,0.5,1,1.5\}$), for the \texttt{syren-new} linear power spectrum.}
\label{fig:matter_consistency}
\end{figure*}

To pin down the numerical differences between the codes we evaluate the $273$ closed triangles of $19$ logarithmic $k$-bins in $[0.01,3]\,h\,{\rm Mpc}^{-1}$ ($\Delta\ln k=0.3$) at $z\in \{0,0.5,1,1.5\}$ on all $1000$ cosmologies of the benchmark set (Section~\ref{sec:cosmologies}). Every code receives the same linear power spectrum table and the $\sigma_8(z)$ that \texttt{bihalofit-JAX} uses in the formula (the requested $\sigma_8$ times the linear growth); \texttt{threepoint\_py} is compared with \texttt{bihalofit-JAX} run on its E\&H spectrum. Figure~\ref{fig:matter_consistency} shows the fractional differences between the codes. We find that T20 and \texttt{threepoint\_py} agree with \texttt{bihalofit-JAX} to better than $0.02\%$ on average (maximum $0.08\%$) at every redshift, and \texttt{fastnc} to $0.02\%$ (maximum $0.2\%$). The CPU build of \texttt{threepoint} agrees to $0.01\%$ at $z=0$ but differs by $0.7\%$ on average at $z=1.5$ (maximum $1.2\%$). We emphasise that this is at the default settings of \texttt{bihalofit-JAX} (Section~\ref{sec:comparison_codes}).

\begin{table*}[htbp]
\centering
\small
\setlength{\tabcolsep}{2pt}
\resizebox{\textwidth}{!}{%
\begin{tabular}{l l c r r r r r r r r r}
\hline\hline
 & & & \multicolumn{4}{c}{\texttt{bihalofit-JAX} [ms]} & \multicolumn{2}{c}{\texttt{threepoint} [ms]} & \texttt{threepoint\_py} [ms] & \texttt{fastnc} [ms] & T20 [ms] \\
\cline{4-7}\cline{8-9}\cline{10-10}\cline{11-11}\cline{12-12}
 & & & \multicolumn{2}{c}{GPU} & \multicolumn{2}{c}{CPU} & GPU & CPU & CPU & CPU & CPU \\
\cline{4-5}\cline{6-7}\cline{8-8}\cline{9-9}\cline{10-10}\cline{11-11}\cline{12-12}
Stat. & Workload & $N_{\rm cosmo}$ & \texttt{fp32} & \texttt{fp64} & \texttt{fp32} & \texttt{fp64} & \texttt{fp64} & \texttt{fp64} & \texttt{fp64} & \texttt{fp64} & \texttt{fp64} \\
\hline
$B_\delta$ & $10{,}160$ & $1$ & $1.44$ & $1.52$ & $0.932$ & $1.41$ & $304$ & $1{,}615$ & $1{,}910$ & $45.8$ & $14.7$ \\
 & (triangles) & $500$ & $1.75$ & $15.3$ & $62.7$ & $146$ & $152{,}030^\S$ & $807{,}459^\S$ & $955{,}239^\S$ & $22{,}910^\S$ & $7{,}353^\S$ \\
$B_\gamma$ & $9{,}799$ & $1$ & $2.48$ & $9.11$ & $25.3$ & $43.7$ & $367$ & $1{,}081$ & $50{,}843$ & $1{,}139$ & -- \\
 & (triangles) & $500$ & $83.4^\ast$ & $2{,}741^\ast$ & $8{,}514$ & $17{,}060$ & $183{,}534^\S$ & $540{,}428^\S$ & $2.54\times10^{7\,\S}$ & $569{,}289^\S$ & -- \\
$C_\ell^{g^2g}$ & $39$ & $1$ & $7.66$ & $163$ & $10{,}125$ & $9{,}870$ & -- & -- & -- & -- & -- \\
 & (multipoles) & $500$ & $1{,}399^\ast$ & $73{,}913^\ast$ & $263{,}036$ & $507{,}975$ & -- & -- & -- & -- & -- \\
$\langle M_{\rm ap}^3\rangle$ & $84$ & $1$ & $5.95$ & $252$ & $1{,}059$ & $1{,}506$ & $67{,}663$ & $148{,}343$ & $71{,}256$ & $20{,}283$ & -- \\
 & (triplets) & $500$ & $2{,}113^\ast$ & $143{,}529^\ast$ & $400{,}378$ & $853{,}978$ & $3.38\times10^{7\,\S}$ & $7.42\times10^{7\,\S}$ & $3.56\times10^{7\,\S}$ & $1.01\times10^{7\,\S}$ & -- \\
\hline
\end{tabular}}

\refstepcounter{table}
\label{tab:projected_speeds}
\begin{minipage}{\textwidth}
\footnotesize
\justifycaption\noindent\textbf{Table \thetable.} Wall time of the statistics considered in this work (matter bispectrum, lensing bispectrum, skew spectra and mass-aperture statistics) across different implementations. These are evaluated for one and $500$ cosmologies, for the devices (CPU/GPU) and floating-point precision available to the respective codes. The matter bispectrum is evaluated at $z=1$. The missing entries reflect the fact that the code does not compute the statistic: only \texttt{bihalofit-JAX} computes skew spectra, and T20 computes only the matter bispectrum. $^\ast$The $500$ cosmologies run in sequential batches to fit in GPU memory: $250$ cosmologies per batch for $B_\gamma$ and the \texttt{float32} skew spectra, $25$ for the \texttt{float64} skew spectra and the \texttt{float32} $\langle M_{\rm ap}^3\rangle$, and $12$ for the \texttt{float64} $\langle M_{\rm ap}^3\rangle$. $^\S$$500$ times the time for one cosmology, as these codes run one cosmology at a time. 
\end{minipage}
\end{table*}

\subsection{Lensing bispectra}\label{sec:timing_bgamma}\label{sec:agreement_projected}

We calculate the weak-lensing bispectrum $B_\gamma$ (without intrinsic alignments) for the auto-correlation of a single source population, on the $9{,}799$ triangles, $\ell_1\leq\ell_2\leq\ell_3\leq\ell_1+\ell_2$, formed from $62$ log-spaced integer multipoles between $\ell=20$ and $1000$. \texttt{threepoint} and \texttt{fastnc} are given the $z=0$ \texttt{syren-new} spectrum as a table together with $\sigma_8$, and scale it to each redshift with a linear growth factor.

For a single cosmology \texttt{bihalofit-JAX} computes $B_\gamma$ in  $\sim 2.5\,\text{ms}$ on GPU and $\sim 25\,\text{ms}$ on CPU in \texttt{float32}, at least $\sim 150$ times faster than the fastest other code, \texttt{threepoint} on GPU, and for $500$ cosmologies at least $\sim 2\times 10^3$ times faster (Table~\ref{tab:projected_speeds}). Note that the GPU is faster than the CPU for the weak-lensing bispectrum considered here, in contrast to the matter bispectrum (Section~\ref{sec:timing_matter}), because the cost of launching the kernel is now subdominant to the cost of the calculation.

\begin{figure*}[htbp]
\centering
\includegraphics[width=\textwidth]{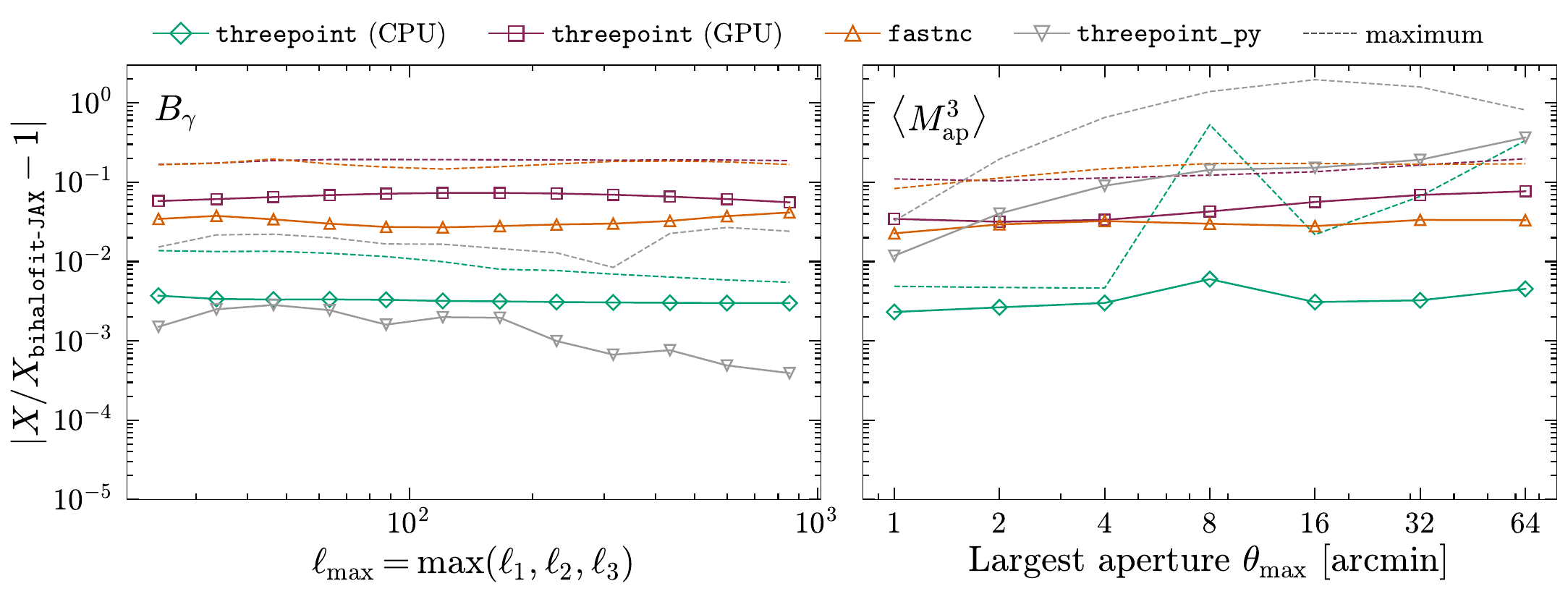}
\caption{Mean (solid) and maximum (dashed) fractional differences compared to \texttt{bihalofit-JAX} (at \texttt{float64}) for the lensing bispectrum $B_\gamma$ against $\ell_{\rm max}=\max(\ell_1,\ell_2,\ell_3)$ (left, in $12$ logarithmic bins of the $9{,}799$ triangles) and for $\langle M_{\rm ap}^3\rangle$ against the largest aperture of the triplet (right, for $84$ triplets). This is calculated over the $30$ cosmologies of Section~\ref{sec:cosmologies}, at the drawn $w_0$ except for the CPU build of \texttt{threepoint}, which is at $w_0=-1$.}
\label{fig:projected_consistency}
\end{figure*}

To assess the differences between the codes, we evaluate the weak-lensing bispectrum for the cosmologies and source distributions described in Section~\ref{sec:cosmologies}, taking the mean and maximum fractional differences over the set of results. We find that the CPU version of \texttt{threepoint} agrees with \texttt{bihalofit-JAX} to $0.3\%$ on average (at most $1.4\%$; Figure~\ref{fig:projected_consistency}, left), and its GPU build differs by $7\%$ on average (at most $19\%$). \texttt{fastnc}, at settings chosen to agree with \texttt{threepoint} to about $1\%$ but with its default line-of-sight redshift bins, agrees to $3\%$ on average (at most $20\%$), and \texttt{threepoint\_py} to $0.13\%$ (at most $2.7\%$). Note that this comparison is for the default accuracy settings of \texttt{bihalofit-JAX}, as described in Section~\ref{sec:comparison_codes}.

\subsection{Skew spectra}\label{sec:timing_skew}

To examine the performance of the skew spectra implementation in \texttt{bihalofit-JAX}, we time the skew spectrum for a representative case of a top-hat filter over $\ell\in[20,400]$ with $39$ output multipoles, with number-count kernels on all three legs (Eq.~\ref{eq:skew_spectrum}) with linear bias $b=1.5$ and a Gaussian redshift distribution, $n_g(z)$, given by 
\begin{equation}
  n_g(z) = \frac{1}{\sqrt{2\pi}\sigma_z}\exp\left(-\frac{(z-\bar z)^2}{2\sigma_z^2}\right),
\end{equation}
with $\bar z=0.5$ and $\sigma_z=0.1$. We keep the other precision settings at their defaults, described in Section~\ref{sec:comparison_codes}. The bispectrum is interpolated on $68$ nodes per side\footnote{$\ln B$ is evaluated on a cube of $68^3$ triangles (nodes linear below $\ell=100$, logarithmic above) and interpolated trilinearly.}, which reproduces the exact double sum to a median error of $0.14\%$ (at most $1\%$, at $\ell=20$, and $0.4\%$ for $\ell\geq 50$). For a single cosmology this takes $\sim 8\,\text{ms}$ on GPU in \texttt{float32} and $\sim 10\,\text{s}$ on a CPU restricted to $24$ cores, and Table~\ref{tab:projected_speeds} also gives the times for $500$ cosmologies. These times grow quickly if one requests many multipoles over a large filter, even on a GPU in \texttt{float32}. No other code we compare with computes skew spectra.

\subsection{Mass-aperture statistics}\label{sec:timing_map3}\label{sec:agreement_map3}

We analyse the performance of the mass-aperture statistics in the various implementations, evaluating Eq.~(\ref{eq:mass_aperture}) for a set of $84$ aperture triplets $\theta_1\leq\theta_2\leq\theta_3$ drawn from $\theta\in\{1,2,4,8,16,32,64\}'$. We use the Crittenden filter and the same lensing tracer on all three legs. \texttt{bihalofit-JAX} integrates all triplets on one diagonal grid of $N_\ell=96$ GL nodes in $\ln\ell_1$, from $\ell_1=1$ to $10/\theta_{\min}$ with $\theta_{\min}$ the smallest aperture in radians, $N_\rho=48$ ratio panels of two nodes each and $N_\varphi=64$ angles. The other settings are held at their defaults, as described in Section~\ref{sec:comparison_codes}.

\texttt{threepoint} integrates each triplet adaptively up to $\ell=10/\theta_{\min}$ of that triplet, with $400$ redshift nodes over $0\le z\le5$ on CPU and the $256$ compiled into its GPU build, and is given the apertures largest first, the order in which it is fastest. \texttt{threepoint\_py} uses its default settings: a bispectrum table of $20$ multipoles and $20$ angles over $1\le\ell\le20000$, and $200$ redshift points in its Limber integral. For \texttt{fastnc} we use $L_{\max}=30$, $M_{\max}=40$, $80$ logarithmically spaced $t_1$ between $0.1'$ and $600'$ and $60$ angles, with its default line-of-sight redshift bins.

For one cosmology, \texttt{bihalofit-JAX} evaluates all $84$ triplets in $6\,{\rm ms}$ on GPU and $1.1\,{\rm s}$ on CPU in \texttt{float32}. On GPU this is $\sim10^4$ times faster than the \texttt{threepoint} CUDA build, demonstrating the advantage of a fixed grid over an adaptive integrator: one can leverage the fact that GPUs are highly efficient at computing large numbers of operations. \texttt{threepoint} integrates each triplet adaptively, so its cost per triplet grows as the triplet is squeezed, whereas the cost of our fixed grid does not depend on the triplet shape. \texttt{threepoint\_py} runs on a single core and repeats a $\sim 11\,{\rm s}$ bispectrum precomputation for every cosmology, whereas our build and compilation are paid once. Even on CPU we find \texttt{bihalofit-JAX} to be $\sim140$ times faster than \texttt{threepoint} and $\sim20$ times faster than \texttt{fastnc} in \texttt{float32}, the fastest of the other codes ($\sim100$ and $\sim13$ times in \texttt{float64}). 

Comparing the numerical precision for the cosmologies and source distributions of Section~\ref{sec:cosmologies}, we find that the CPU build of \texttt{threepoint} agrees to $0.4\%$ on average with \texttt{bihalofit-JAX} (Figure~\ref{fig:projected_consistency}, right). For $10$ of the $2{,}520$ inputs its adaptive integrator returns a wrong value without warning, off by $5$--$53\%$, and these set its maximum. The GPU build of \texttt{threepoint} differs from \texttt{bihalofit-JAX} by $6\%$ on average (at most $20\%$) and returns exact zeros for $10$ of the $84$ triplets (we exclude these from the comparison). \texttt{fastnc} agrees to $3\%$ on average (at most $17\%$). \texttt{threepoint\_py}, at the default settings its own examples use, differs by $22\%$ on average (at most $200\%$).

\section{Gradient-Based Inference}\label{sec:performance}

\subsection{Likelihood Construction}

This section is intended primarily as a test of the use of \texttt{bihalofit-JAX} in inference.  We consider a simplified scenario to test our code against the original T20 code. We run a mock cosmological parameter
inference on the weak lensing bispectrum for an autocorrelation of the Smail-type source distribution of Eq.~(\ref{eq:Smail}) with $z_0=0.3$, $\alpha=2$, and $\beta=1$. We evaluate the Limber integral from $z=0$ to $z=3.5$, with $n_z=200$ linearly spaced redshifts in the T20 evaluation and the default $N_\chi=160$ comoving-distance nodes over $10\le\chi\le5800\,h^{-1}{\rm Mpc}$ in \texttt{bihalofit-JAX} (the other precision settings are held at their defaults: $N_R=15$, $N_z^{\rm bg}=2000$ uniformly spaced redshifts up to $z=10$ for $\chi(z)$, and $N_k=600$ log-spaced points over $10^{-4}\le k\le50\,h\,{\rm Mpc}^{-1}$). We create a mock data vector
from the BiHalofit prediction using T20, evaluated at a fiducial cosmology of $\Omega_{\rm m} = 0.3156$, $\Omega_{\rm b}=0.0492$, $h=0.6727$, $n_s=0.9645$, $\sigma_8=0.831$, $w_0=-1$, $w_a=0$, and $M_\nu=0.0\,{\rm eV}$. We then project it onto the weak-lensing bispectrum $B_\gamma$ with the Limber integral of Eq.~(\ref{eq:projected_bispectra}), using the lensing kernel of Eq.~(\ref{eq:wlensing}) on all three legs (Section~\ref{sec:bkappa}). The same $\chi(z)$ and kernel are used as in \texttt{bihalofit-JAX}, so that the only difference between the mock and the model is the BiHalofit implementation: T20 for $B_\delta$ in the data vector and \texttt{bihalofit-JAX} in the model.

Both \texttt{bihalofit-JAX} and T20 use the \texttt{syren-new} linear power
spectrum backend, so that the comparison isolates the
BiHalofit kernel implementation and the sampler rather than also
mixing in a linear power spectrum discrepancy.

We assume for simplicity a diagonal, Gaussian covariance in the flat-sky limit for the
binned $B_\gamma$ data vector \citep{Takada_2004, Joachimi_2009, Kayo_2012},
\begin{equation}
  \mathrm{Cov}_{ii} = s_{123}\,\frac{\tilde{C}(\ell_{1,i})\,\tilde{C}(\ell_{2,i})\,
  \tilde{C}(\ell_{3,i})}{N_{\rm tri}(\ell_{1,i},\ell_{2,i},\ell_{3,i})},
\end{equation}
for each triangle bin $i$, where $\tilde{C}(\ell)=C_\gamma(\ell)+\sigma_\epsilon^2/\bar{n}$ is the observed lensing power spectrum, the sum of the signal and the shape noise. In the Limber approximation the signal is
\begin{equation}\label{eq:cgamma}
  C_\gamma(\ell)=\int\frac{{\rm d}\chi}{\chi^2}\,W_{\rm lensing}^2(\chi)\,P_\delta\!\left(\frac{\ell+\frac12}{\chi};z(\chi)\right),
\end{equation}
with $W_{\rm lensing}$ the lensing kernel of Eq.~(\ref{eq:wlensing}) and $P_\delta$ the nonlinear matter power spectrum. The factor $s_{123}\in\{1,2,6\}$ is equal to $1$ if all the $\ell$'s are different, $2$ if two are repeated and $6$ if all three are the same, and $N_{\rm tri}$ is the number of Fourier-mode triangles falling in that bin triplet,
\begin{equation}
\begin{split}
  N_{\rm tri}(\ell_1,\ell_2,\ell_3) &= \frac{\Omega_{\rm s}}{(2\pi)^3}
  \int_{\ell_1}\!\!\ell_1'\mathrm{d}\ell_1' \int_{\ell_2}\!\!\ell_2'\mathrm{d}\ell_2'\\
  &\quad \times \int_0^{2\pi}\!\!\mathrm{d}\varphi\;
  \mathbf{1}\!\left[\lvert\vec{\ell}_1'+\vec{\ell}_2'\rvert \in \ell_3\text{-bin}\right],
\end{split}
\end{equation}
with $\Omega_{\rm s}$ the survey solid angle and the two inner integrals over each bin's width. We evaluate $C_\gamma$ with the Core Cosmology Library (\texttt{CCL}) \citep{Chisari_2019}, using its Halofit power spectrum, at the fiducial cosmology, and $N_{\rm tri}$ by direct quadrature over $(\ell_1',\ell_2',\varphi)$. We then add a shape noise of $\sigma_\epsilon=0.26$ and a source density of $\bar n=27\,{\rm arcmin}^{-2}$, and assume a sky fraction $f_{\rm sky}=0.4$ ($\Omega_{\rm s}=4\pi f_{\rm sky}$).

The data vector consists of $B_\gamma$ in the $126$ closed triangles of $12$ logarithmically spaced multipole bins between $\ell=50$ and $\ell=2000$, evaluated at the geometric bin centres ($\ell_1\le\ell_2\le\ell_3$ with $\ell_3<\ell_1+\ell_2$). Sampling proceeds on the Gaussian log-likelihood
\begin{equation}
\begin{split}
  -2\ln L(\theta) &= \sum_i
  \frac{1}{\mathrm{Cov}_{ii}}\Big[B_\gamma^{\rm truth}(\ell_{1,i},\ell_{2,i},\ell_{3,i})\\
  &\quad - B_\gamma(\ell_{1,i},\ell_{2,i},\ell_{3,i};\theta)\Big]^2,
\end{split}
\end{equation}
with $B_\gamma^{\rm truth}$ the mock data vector above and
$B_\gamma(\theta)$ the model under consideration evaluated at the sampled
parameters $\theta$. 

\subsection{Results}

\begin{figure}[htbp]
\centering
\includegraphics[width=\columnwidth]{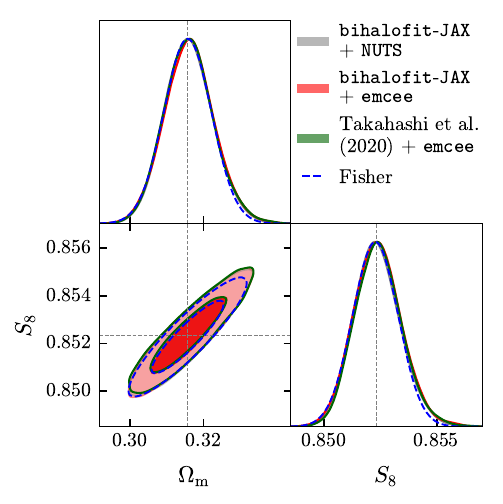}
\caption{Posterior constraints on $(\Omega_{\rm m}, S_8)$, with $S_8=\sigma_8(\Omega_{\rm m}/0.3)^{0.5}$, recovered from the mock
lensing bispectrum data vector with $h$ and $n_s$ fixed. \texttt{bihalofit-JAX} with \texttt{NUTS} is shown in grey,
\texttt{bihalofit-JAX} with \texttt{emcee} in red, T20 with \texttt{emcee}
in dark green, and the Fisher forecast is shown in blue. The dotted lines
mark the injected cosmology. All four agree and recover the injected cosmology to better than $0.1\sigma$.}
\label{fig:mcmc_triangle}
\end{figure}

We sample the two cosmological parameters that
are important for the lensing-bispectrum, $\Omega_{\rm m}$ and $\sigma_8$ (the mock has $M_\nu=0$, so $\Omega_{\rm m}=\Omega_{\rm cb}$), with $h$ and $n_s$ fixed at their fiducial values\footnote{Sampling over $\Omega_{\rm m}$ is equivalent to sampling over the physical density $\Omega_{\rm m}h^2$ since $h$ is fixed.}. The priors are uniform in $\ln\Omega_{\rm m}$ and $\ln\sigma_8$ over $\Omega_{\rm m}\in[0.24,0.40]$, the validity range of the \texttt{syren-new} linear power spectrum (Table~\ref{tab:calibration_ranges}), and $\sigma_8\in[0.50,1.30]$, which takes the lower bound of BiHalofit's calibrated range. We use two independent samplers on this likelihood: the
No-U-Turn Sampler (NUTS) via \texttt{numpyro} \citep{numpyro2019}, using
exact gradients of the likelihood from \texttt{jax.grad}, and an
affine-invariant ensemble sampler via \texttt{emcee} \citep{emcee2013}
as a gradient-free comparison. Additionally, we perform a Fisher-matrix forecast \citep{fisher_original, Tegmark_1997}, 
\begin{equation}
  F = \left(\frac{\partial B_\gamma}{\partial\theta}\right)^\top
C^{-1} \left(\frac{\partial B_\gamma}{\partial\theta}\right),
\end{equation}  
where $\theta$ is the vector of parameters and $C$ the covariance. We 
compute the derivatives via
\texttt{jax.jacfwd}. Finally, we also
run \texttt{emcee} directly on the original, non-differentiable T20 code. Note that we use \texttt{float64} for \texttt{bihalofit-JAX} paired with NUTS since we found gradients to be more stable at this precision (it is run in \texttt{float32} with \texttt{emcee}). All four recover
the fiducial cosmology within their statistical uncertainty, and their
marginal uncertainties agree closely with one another, as shown in
Figure~\ref{fig:mcmc_triangle}.

The samplers work in the logarithms of the parameters, $\theta=(\ln\Omega_{\rm m},\ln\sigma_8)$. NUTS runs $8$ chains with $600$ warm-up steps and a target acceptance probability of $0.8$, drawing in chunks of $500$ until the chains converge. Both \texttt{emcee} runs, with \texttt{bihalofit-JAX} and with T20, use $32$ walkers for the same $1000$ steps. For each parameter we calculate the effective sample size, the number of statistically independent samples in the chains \citep{Gelman2013},
\begin{equation}
  {\rm ESS}\equiv\frac{N_{\rm c}N_{\rm s}}{\tau},
\end{equation}
where $N_{\rm c}$ is the number of chains (or walkers), $N_{\rm s}$ the number of post-warm-up steps per chain, and $\tau$ the integrated autocorrelation time. We report the minimum over the two parameters.

\begin{table*}[t]
\centering
\small
\setlength{\tabcolsep}{5pt}
\begin{tabular}{l l c c c c c}
\hline\hline
Code & Sampler & Samples & Wall time [s] & Min.\ ESS & ESS/s & Relative ESS/s \\
\hline
\texttt{bihalofit-JAX} & \texttt{emcee} & $32\times1000$ & $14.0$ & $3712$ & $265$ & $1$ \\
\texttt{bihalofit-JAX} & NUTS & $8\times500$ & $95.4$ & $3277$ & $34.3$ & $0.13$ \\
T20 & \texttt{emcee} & $32\times1000$ & $5335$ & $3292$ & $0.617$ & $0.0023$ \\
\hline
\end{tabular}

\refstepcounter{table}
\label{tab:ess}
\begin{minipage}{\textwidth}
\footnotesize
\justifycaption\noindent\textbf{Table \thetable.} Sampling efficiency for the mock inference of Section~\ref{sec:performance}. The samples column gives the number of chains (walkers) times the samples in each; both \texttt{emcee} runs take $1000$ steps. The minimum ESS is over $(\Omega_{\rm m},\sigma_8)$, and the relative ESS/s is normalised to \texttt{bihalofit-JAX} with \texttt{emcee}. Wall times include compilation and, for NUTS, warm-up. The \texttt{bihalofit-JAX} runs use the GPU of Section~\ref{sec:comparison_codes} (\texttt{emcee} in \texttt{float32}, NUTS in \texttt{float64}). For T20, the $32$ CPU processes are run in \texttt{float64}.
\end{minipage}
\end{table*}

The lensing bispectrum alone leaves a strong degeneracy between
$\sigma_8$ and $\Omega_{\rm m}$, consistent with the well-known
bispectrum degeneracy direction. This causes problems for NUTS, so we apply a
standard covariance-whitening (Cholesky-decomposition) approach \citep{LewisBridle2002,Brooks_2011,Lewis_2013}, with the
covariance obtained analytically from the Fisher matrix evaluated at the fiducial cosmology
\begin{equation}
  \theta = \theta_0 + L y, \qquad L L^\top = F^{-1}(\theta_0),
\end{equation}
where $\theta_0$ is the fiducial
cosmology, and $L$ the Cholesky factor. 

Overall we find that
\texttt{bihalofit-JAX} with NUTS draws $8\times500$ samples (minimum ESS $3277$) in $95.4\,\rm{s}$, and \texttt{bihalofit-JAX} with \texttt{emcee} draws $32\times1000$ steps (minimum ESS $3712$) in $14.0\,\rm{s}$. This is shown in Table~\ref{tab:ess}. We find that \texttt{emcee} is both faster in
wall-clock time and $\sim$8$\times$ more efficient
in ESS/s on this low-dimensional posterior because of both the increased cost of computing derivatives and the precision required to compute them (we use \texttt{float64} with NUTS). T20 with \texttt{emcee}, run for the same $1000$ steps with $32$ processes, needs $5335$\,s to reach a comparable ESS, which is $\sim$430$\times$ slower in ESS/s than \texttt{bihalofit-JAX} with \texttt{emcee}
despite parallelising the model calls across $32$ processes. This is because the \texttt{bihalofit-JAX} model call is vectorised and GPU-optimised. Independent comparisons of gradient-based and gradient-free sampling on cosmological likelihoods report a comparable finding \citep{SwiftCl2025, mootoovaloo2024assessmentgradientbasedsamplersstandard, Ruiz_Zapatero_2024}: once the cost of the gradient is included, the advantage of gradient-based samplers is modest, or absent, at low to moderate dimensionality, and it is expected to reassert itself at higher dimensionality, where ensemble-sampler acceptance rates degrade sharply.

\section{Discussion and Conclusion}\label{sec:conclusion}

In this paper we have presented a new open source code, \texttt{bihalofit-JAX}, for the matter bispectrum and related projected statistics such as projected bispectra, skew spectra and aperture statistics. Bispectrum-related statistics are notoriously more complicated and expensive to compute, leading to intricate approaches such as neural network emulators which are designed to replace the numerical calculations. We have shown here that these emulators are in fact not needed for these types of problems, where the calculations can be made far more efficient and optimised for GPUs. We have achieved this goal by utilising fixed grids which can be evaluated extremely efficiently on GPUs, and designing carefully around the differentiability of the code (care has to be taken beyond just re-writing codes in \texttt{JAX}). This can be seen very clearly by comparing our code to other implementations, where our code computes the matter bispectrum for $500$ different cosmologies in $\sim 2\,\text{ms}$, around $4\times 10^3$ times faster than the next fastest implementation and $5\times 10^5$ times faster than the slowest. 

The BiHalofit formula has been re-implemented many times in various public and private codes, and we aim to provide a better validated version for the community to use. This code will be useful for many types of cosmological analyses, such as the construction of emulators for the matter bispectrum (where emulators are more relevant due to expensive simulations required to generate accurate nonlinear predictions) which will greatly benefit from simpler, fast models that can be used to train against as a baseline.  \texttt{bihalofit-JAX} could present useful progress in analytical understanding of the constraining power and degeneracy-breaking potential of different bispectrum-based statistics applied to stage-IV surveys. 

Several improvements remain possible. The main limitation of our approach is the use of fixed integration grids: evaluating the formulae far outside the regime we tested could silently degrade the precision. We validated the precision settings over Latin hypercubes of inputs in Appendix~\ref{app:precision} (for the aperture statistics for lensing only, the case the other codes can check), but a more robust implementation would refine the grids adaptively, in a way that remains efficient on GPUs and limits host-to-device transfers. Whether such a scheme works reliably across the full parameter space would need to be checked carefully. One other small point worth making is that the code could be made more memory efficient by splitting the calculation into chunks. This would be important for the skew spectra, where the large double sum over multipoles is memory intensive for GPU work. 

The code also computes one type of statistic at a time, for a single redshift distribution in the case of projected statistics. Many analyses combine several tracers in one data vector, and we expect the code could be batched more efficiently over tracers, which we plan to implement in a future release. 

The computation could be accelerated further by emulating the nonlinear scale, $k_{\text{NL}}$, and the effective spectral index, $n_{\text{eff}}$, of Eqs.~(\ref{eq:sigma2_kNL}) and (\ref{eq:neff}). Symbolic regression, as in \citet{Bartlett_2024_halofit}, can give simple, differentiable and extremely fast approximations to these intermediate quantities. We do not adopt this here because the code is already fast and this way keeps numerical control. Emulating the full BiHalofit formula would be much harder: it would need a large network and training set, and would likely be slower (because of the number of nodes) and less accurate than emulating the intermediate quantities. A flexible code is also easier to modify than an emulator is to retrain, for example to inject a nonlinear power spectrum in place of the linear one. Related to this, it might be worth also providing utility in a future version to run the linear power spectrum backend directly with \texttt{CLASS} or \texttt{CAMB}, if speed is less of a concern but higher fidelity results are required, e.g. over a range of cosmologies outside the training range of the emulated backends. 
\section{Availability}
\texttt{bihalofit-JAX} is open source under the MIT licence and available at \url{https://github.com/AlexRoskill/bihalofit-jax}. It is available on PyPI and can be installed with \texttt{pip install bihalofit-jax}, with the optional \texttt{baccoemu} backend enabled by installing \texttt{bihalofit-jax[baccoemu]}. The repository contains the documentation and tutorial notebooks for each statistic in this paper. \noindent\textit{Software:} \texttt{JAX} \citep{jax2018github},
\texttt{NumPy} \citep{numpy2020}, \texttt{baccoemu} \citep{Angulo2021},
\texttt{numpyro} \citep{numpyro2019}, \texttt{emcee} \citep{emcee2013},
\texttt{GetDist} \citep{Lewis2019GetDist}, \texttt{CCL} \citep{Chisari_2019}, \texttt{syren-new} \citep{Sui_2025} and
\texttt{pyshtools} \citep{https://doi.org/10.1029/2018GC007529}. The codes compared in Section~\ref{sec:benchmarks} are used as described in Section~\ref{sec:implementations}, at the following versions: the BiHalofit code of \citet{Takahashi2020} as released on the authors' website, \texttt{fastnc} v1.2.3, \texttt{threepoint\_py} v1.0.0 \citep{2026zndo..19072610L}, and \texttt{threepoint} at commit \texttt{08de964} for the CPU build and \texttt{33c389b} for the GPU build.

\acknowledgments
\section*{Acknowledgements}
We thank Pedro Ferreira for encouraging us to write this paper, as well as Deaglan Bartlett, Christopher Everett and David Alonso, for interesting discussions. We thank Andrina Nicola and Léa Harscouet for making available their filter-squared bispectrum code. AR acknowledges support from STFC. MZ acknowledges that the project that gave rise to these results received the support of a fellowship from the ‘La Caixa’ Foundation (ID 100010434);
the fellowship code is LCF/BQ/PI25/12100027. We acknowledge the use of the DIPC Supercomputing Center and thank Ra\'ul Angulo for the use of the new GPU. The software 
development for this work was assisted by
Claude Code, for which we take full responsibility. The paper was written by us with minimal assistance from Large Language Models.

\appendix

\section{Fitting-Formula Coefficients}\label{app:coeffs}

\subsection{BiHalofit Fitting-Formula Coefficients}\label{app:bhf_coeffs}

Using $n_{\rm eff}$ for the effective spectral index, $\sigma_8$, $\Omega_{\rm m}$,
and the triangle-shape parameters $r_1,r_2$ of Section~\ref{sec:model}, the
coefficients of Eqs.~(\ref{eq:B3h})--(\ref{eq:Pe}) are power laws,
most conveniently written in log form, matching the presentation in \citet{Takahashi2020}. The fitting parameters of the $1$-halo
term, $T(q)$, are
\begin{align}
  \log_{10} a_n &= -2.167 - 2.944\log_{10}\sigma_8
    - 1.106\left(\log_{10}\sigma_8\right)^2 \notag \\
    &\quad{} - 2.865\left(\log_{10}\sigma_8\right)^3 - 0.310\,r_1^{\gamma_n}, \\
  \log_{10} b_n &= -3.428 - 2.681\log_{10}\sigma_8
    + 1.624\left(\log_{10}\sigma_8\right)^2 \notag \\
    &\quad{} - 0.095\left(\log_{10}\sigma_8\right)^3, \\
  \log_{10} c_n &= 0.159 - 1.107\,n_{\rm eff}, \\
  \log_{10}\alpha_n &= \min\!\left[-4.348 - 3.006\,n_{\rm eff} - 0.5745\,n_{\rm eff}^2
    + 10^{\,-0.9+0.2\,n_{\rm eff}}\,r_2^2,\ \log_{10}\!\left(1-\tfrac{2}{3}n_s\right)\right], \label{eq:alpha_n} \\
  \log_{10}\beta_n &= -1.731 - 2.845\,n_{\rm eff} - 1.4995\,n_{\rm eff}^2 \notag \\
    &\quad{} - 0.2811\,n_{\rm eff}^3 + 0.007\,r_2, \\
  \log_{10}\gamma_n &= 0.182 + 0.57\,n_{\rm eff}, \\
  \intertext{and those of the $3$-halo term, $B_{\rm 3h}$, are}
  \log_{10} d_n &= -0.483 + 0.892\log_{10}\sigma_8 - 0.086\,\Omega_{\rm m}, \\
  \log_{10} e_n &= -0.632 + 0.646\,n_{\rm eff}, \\
  \log_{10} f_n &= -10.533 - 16.838\,n_{\rm eff} - 9.3048\,n_{\rm eff}^2
    - 1.8263\,n_{\rm eff}^3, \\
  \log_{10} g_n &= 2.787 + 2.405\,n_{\rm eff} + 0.4577\,n_{\rm eff}^2, \\
  \log_{10} h_n &= -1.118 - 0.394\,n_{\rm eff}, \\
  \log_{10} m_n &= -2.605 - 2.434\log_{10}\sigma_8
    + 5.71\left(\log_{10}\sigma_8\right)^2, \\
  \log_{10} n_n &= -4.468 - 3.08\log_{10}\sigma_8
    + 1.035\left(\log_{10}\sigma_8\right)^2, \\
  \log_{10} \mu_n &= 15.312 + 22.977\,n_{\rm eff} + 10.9579\,n_{\rm eff}^2 \notag \\
    &\quad{} + 1.6586\,n_{\rm eff}^3, \\
  \log_{10} \nu_n &= 1.347 + 1.246\,n_{\rm eff} + 0.4525\,n_{\rm eff}^2, \\
  \log_{10} p_n &= 0.071 - 0.433\,n_{\rm eff}.
\end{align}
\texttt{bihalofit-JAX} evaluates the minimisation of Eq.~(\ref{eq:alpha_n})
with a $C^1$ smooth-min, to keep derivatives continuous.

\subsection{Baryonic-Feedback Correction Coefficients}\label{app:baryon}

Eq.~(\ref{eq:baryon_leg}) gives the  baryon-feedback ratio, where \citet{Takahashi2020} fits all of these as
functions of the scale factor $a$, apart from the constant exponents $\alpha_0$ and $\alpha_2$, 
\begin{align}
  A_0(a) &= \begin{cases} 0.068\,(a-0.5)^{0.47}, & a > 0.5 \\
                          0, & a \le 0.5 \end{cases}, \\
  \mu_0(a) &= 0.018\,a + 0.837\,a^2, \\
  \sigma_0(a) &= 0.881\,\mu_0(a), \\
  \alpha_0 &= 2.346, \\[4pt]
  A_1(a) &= \begin{cases} 1.052\,(a-0.2)^{1.41}, & a > 0.2 \\
                          0, & a \le 0.2 \end{cases}, \\
  \mu_1(a) &= \left|\,0.172 + 3.048\,a - 0.675\,a^2\,\right|, \\
  \sigma_1(a) &= (0.494 - 0.039\,a)\,\mu_1(a), \\[4pt]
  k_s(a) &= 29.90 - 38.73\,a + 24.30\,a^2 \quad [h\,{\rm Mpc}^{-1}], \\
  \alpha_2 &= 2.25, \\
  \beta_2(a) &= \frac{0.563}{\alpha_2}\,
    \left[\left(\frac{a}{0.060}\right)^{0.02} + 1\right]^{-1}.
\end{align}
As noted in
Section~\ref{sec:model}, the fit is calibrated only to $z=5$.

\section{Precision settings}\label{app:precision}

This appendix tests the precision settings of the three numerical components of the pipeline: the BiHalofit kernel (Section~\ref{sec:kernel}), the projection integral (Section~\ref{sec:bkappa}) and the aperture-mass integral (Section~\ref{sec:mapn}).

\subsection{BiHalofit kernel}\label{app:bdelta}

\begin{figure*}[htbp]
  \centering
  \includegraphics[width=\textwidth]{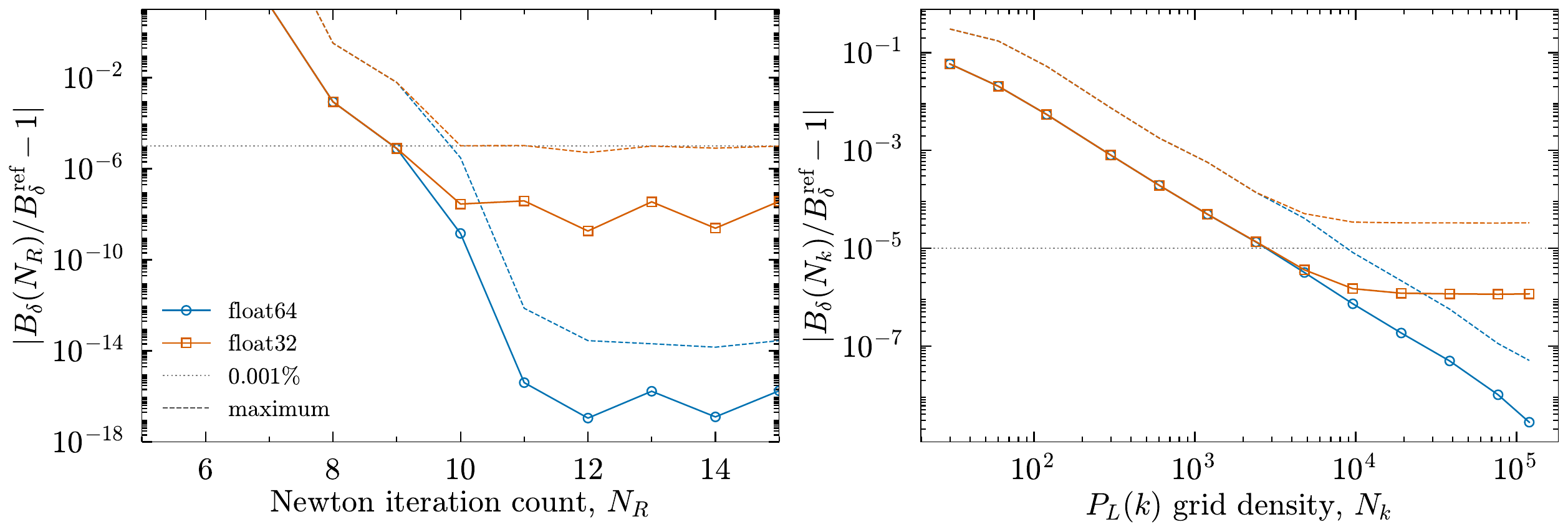}
  \caption{Relative error of the Bihalofit matter bispectrum against Newton iteration count $N_R$ (left) and linear power spectrum grid density
  $N_k$ (right). Solid lines with markers are the mean and dashed lines the maximum over the input cosmologies, triangles and redshifts.}
  \label{fig:bdelta_NR_convergence}
\end{figure*}

The BiHalofit formula in \texttt{bihalofit-JAX} is evaluated on a linear power spectrum grid of $N_k$ points, and its $k_{\rm NL}$ solve takes $N_R$ Newton iterations (Section~\ref{sec:kernel}). We test here the convergence of these precision parameters for the \texttt{syren-new} linear power spectrum backend, since \texttt{baccoemu} has a fixed $N_k=600$. We find similar results for the E\&H backend. To test the convergence of these parameters over a robust range of inputs we generate $1000$ points of a Latin hypercube of cosmologies in the intersection of the training ranges of  \texttt{syren-new} and BiHalofit, shown in the last column of Table~\ref{tab:calibration_ranges}. We evaluate BiHalofit at $z=0,0.5,1,1.5$ on all $542$ closed triangles of $25$ logarithmic $k$-bins in $[0.01,3]\,h\,{\rm Mpc}^{-1}$ ($\Delta\ln k\simeq 0.23$). Figure~\ref{fig:bdelta_NR_convergence} shows the relative error of $B_\delta$ against a reference with $N_R=100$ and $N_k=150{,}000$, varying one setting at a time (every $k$ grid is log-spaced over $10^{-4}\le k\le50\,h\,{\rm Mpc}^{-1}$), in \texttt{float32} and \texttt{float64}\footnote{The optional baryonic-suppression term (Section~\ref{sec:model}) is a closed-form factor in $k_1,k_2,k_3$ and $a$ alone, independent of $N_R$, $N_k$ and the backend, so it does not affect these settings.}. The error falls rapidly with $N_R$ and reaches machine precision at $N_R\sim 12$, and at $N_k=600$ the grid error is $1.9\times10^{-4}$ on average and at most $1.8\times10^{-3}$. 

\subsection{Projection integral}\label{app:bkappa}
\begin{figure*}[htbp]
  \centering
  \includegraphics[width=\textwidth]{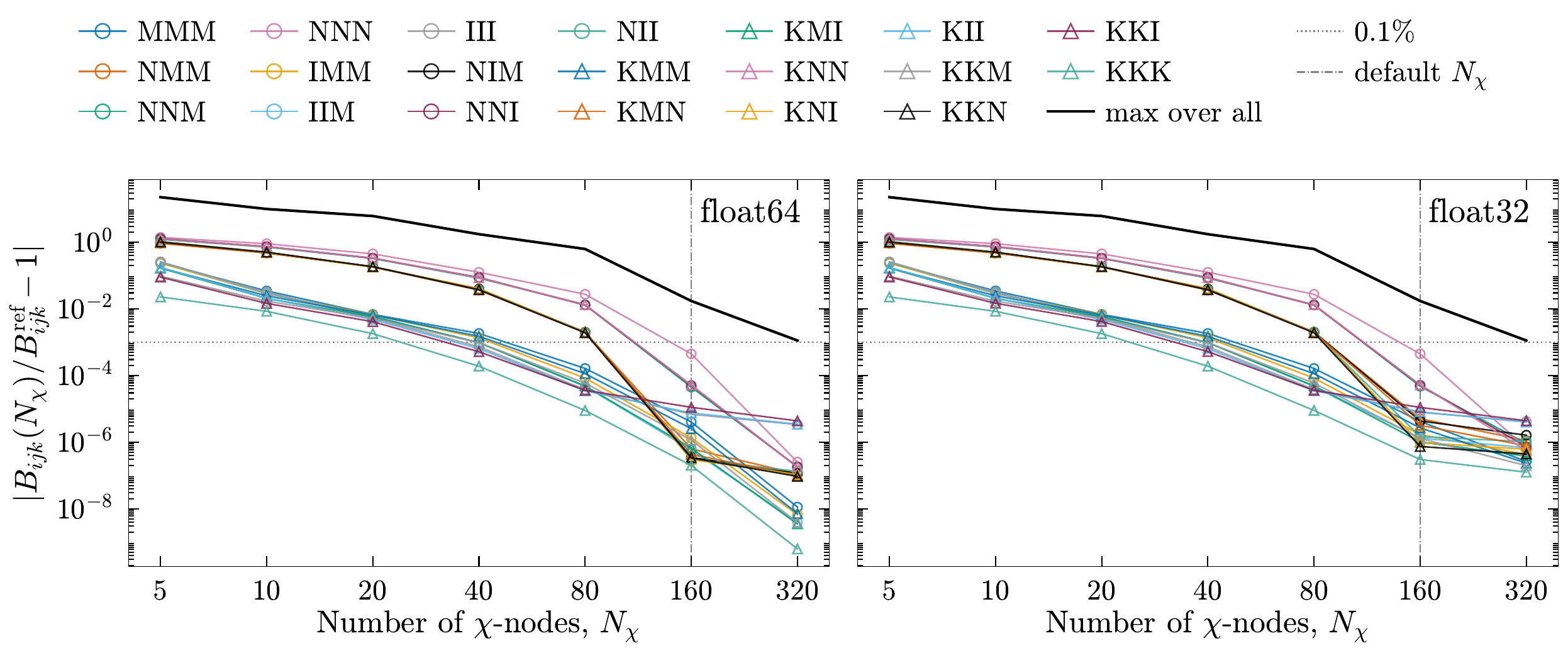}
  \caption{The error in the projected bispectrum $B_{ijk}(\ell_1,\ell_2,\ell_3)$ for a given number of
  GL nodes, $N_\chi$, used in the numerical evaluation of the integral in
  Eq.~(\ref{eq:projected_bispectra}), in \texttt{float64} (left) and \texttt{float32} (right). We denote for brevity: M for lensing, N for counts, I for intrinsic alignment, K for CMB lensing, one letter per leg. We use the
  \texttt{syren-new} linear power spectrum.  Lines show the mean over the $200$-point Latin Hypercube of 
  cosmologies, redshift distributions and triangles, (details elucidated in the text). The black line is the maximum over all
  combinations, cosmologies and triangles, the dotted line marks $0.1\%$ and the dash-dotted line the default
  $N_\chi=160$.}
  \label{fig:bkappa_nchi_convergence}
\end{figure*}

The accuracy of the Limber integral, Eq.~(\ref{eq:projected_bispectra}), is controlled by the number of GL nodes, $N_\chi$. In the case of the projected bispectra one could consider different tracer combinations which might have different convergence properties.  We test all twenty tracer combinations considered in this work, with the \texttt{syren-new} backend, in \texttt{float64} and \texttt{float32}, against $N_\chi=640$ in the same precision. In all cases $N_\chi$ places its GL nodes over $10\le\chi\le5800\,h^{-1}{\rm Mpc}$. The other settings are converged ($N_R=100$, $N_z^{\rm bg}=20000$ uniformly spaced redshifts up to $z=10$ for $\chi(z)$, and $N_k=76800$ log-spaced points over the range $10^{-4}\le k\le50\,h\,{\rm Mpc}^{-1}$). This is shown in Figure~\ref{fig:bkappa_nchi_convergence}. Each combination is evaluated on the $1{,}258$ closed triangles, formed from $30$ log-spaced integer multipoles between $\ell=20$ and $1000$. We evaluate a $200$ point Latin Hypercube over cosmology and redshift distributions. The cosmologies are drawn from the last column of Table~\ref{tab:calibration_ranges}. Lensing and IA legs use Smail source distributions, Eq.~(\ref{eq:Smail}), with $z_0\in[0.15,0.60]$, $\alpha\in[1,3]$ and $\beta\in[0.8,2]$, and the IA redshift index of Eq.~(\ref{eq:wia}) is drawn from $\eta\in[-1,1]$. Counts legs use Gaussian lens distributions with mean $\bar z\in[0.2,0.7]$ and width $\sigma_z\in[0.01,0.10]\,(1+\bar z)$. All $n(z)$ are tabulated on $5000$ linearly spaced points over $0\le z\le5$.

The error is set by the number of counts legs, through the width of the lens sample relative to the node spacing. At $N_\chi=160$ the worst combination has a mean error of $0.05\%$ ($0.001\%$ without a counts leg) and a maximum over all triangles of $2\%$; at $N_\chi=80$ these rise to $2.7\%$ and $60\%$. The \texttt{float32} error reaches its floor at $\sim10^{-4}\%$. 

\subsection{Mass-aperture quadrature}\label{app:map3}

\begin{figure*}[htbp]
\centering
\includegraphics[width=\textwidth]{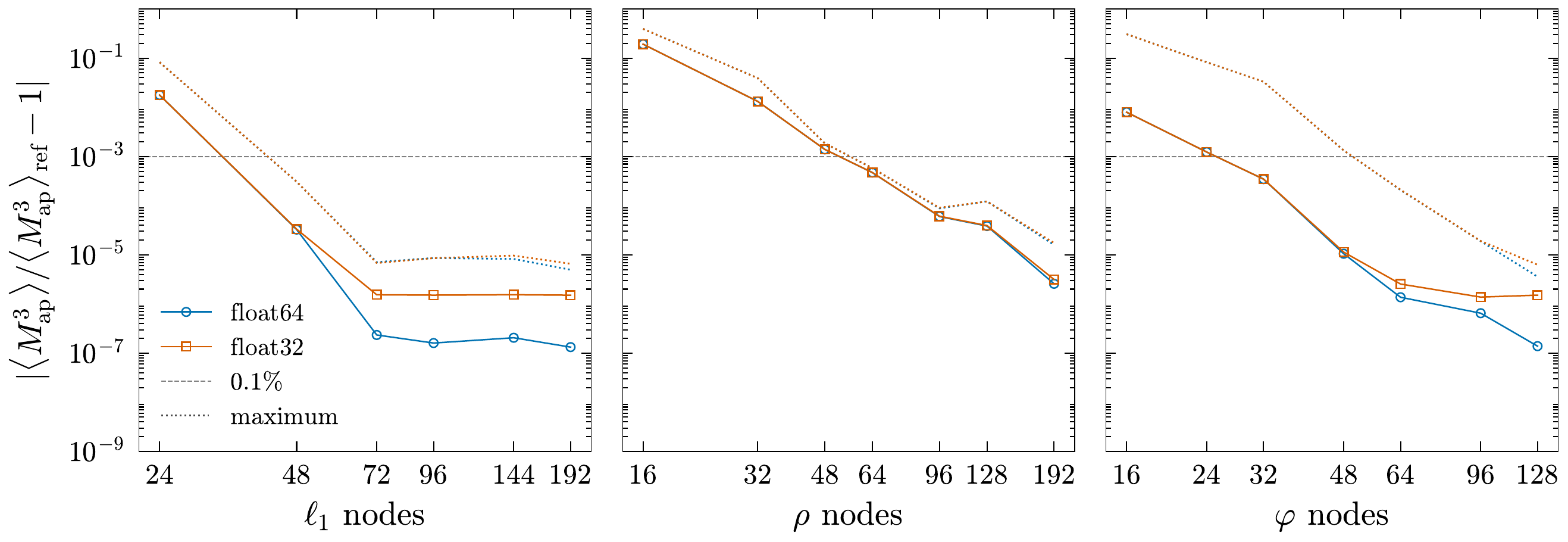}
\caption{Convergence of the $\langle M_{\rm ap}^3\rangle$ quadrature in $N_\ell$ (left), $N_\rho$ (centre; $2N_\rho$ nodes) and $N_\varphi$ (right). Each axis is varied with the other two at their defaults and compared with a \texttt{float64} reference finer on that axis only, using the \texttt{syren-new} linear power spectrum. Blue: \texttt{float64}; orange: \texttt{float32}. Solid lines: mean over triplets, cosmologies and source distributions; dotted: maximum; dashed: $0.1\%$.}
\label{fig:map3_convergence}
\end{figure*}

Figure~\ref{fig:map3_convergence} shows the convergence of the diagonal-grid quadrature of Section~\ref{sec:mapn} for the mass-aperture statistics $\langle M_{\rm ap}^3\rangle$, over the $84$ aperture triplets $\theta_1\leq\theta_2\leq\theta_3$ drawn from $\theta\in\{1,2,4,8,16,32,64\}'$. The grid has $N_\ell$ GL nodes in $\ln\ell_1$ over $1\le\ell_1\le10/\theta_{\min}$ ($\ell_1\le34{,}377$ for $\theta_{\min}=1'$), $N_\rho$ panels of two GL nodes each in $|\ln\rho|$, spaced geometrically from $|\ln\rho|=3\times10^{-3}$ to $\ln(10/\theta_{\min})$ and mirrored about $\rho=1$ (so $2N_\rho$ nodes), and $N_\varphi$ GL nodes over $0\le\varphi\le\pi$. 

Each plot of Figure~\ref{fig:map3_convergence} varies one of $N_\ell$, $N_\rho$ and $N_\varphi$ in \texttt{float32} and \texttt{float64}, with the other two at their defaults ($N_\ell=96$, $N_\rho=48$, $N_\varphi=64$), and compares the results with a \texttt{float64} reference that is finer on the varied axis only ($N_\ell=288$, $N_\rho=128$ or $N_\varphi=192$). Every run, including these references, shares the remaining settings: $N_\chi=160$ GL nodes over $10\le\chi\le5800\,h^{-1}{\rm Mpc}$, $N_R=15$, $N_z^{\rm bg}=2000$ uniformly spaced redshifts up to $z=10$, $N_k=1200$ log-spaced points over $10^{-4}\le k\le10^{4}\,h\,{\rm Mpc}^{-1}$, and $n(z)$ tabulated on $1024$ linearly spaced points over $0\le z\le5$. This is the same in both test and reference, so that the curve isolates the parameter under consideration. The convergence is tested by computing an average over $200$ cosmologies, drawn from the last column of
Table~\ref{tab:calibration_ranges}, each paired with one of ten Smail source distributions ($z_0\in[0.15,0.60]$,
$\alpha\in[1,3]$, $\beta\in[0.8,2]$).

\clearpage
\section{Linear power spectrum backend benchmarks}\label{app:backends}

\subsection{Relative accuracy}

\begin{figure*}[htbp]
\centering
\includegraphics[width=\textwidth]{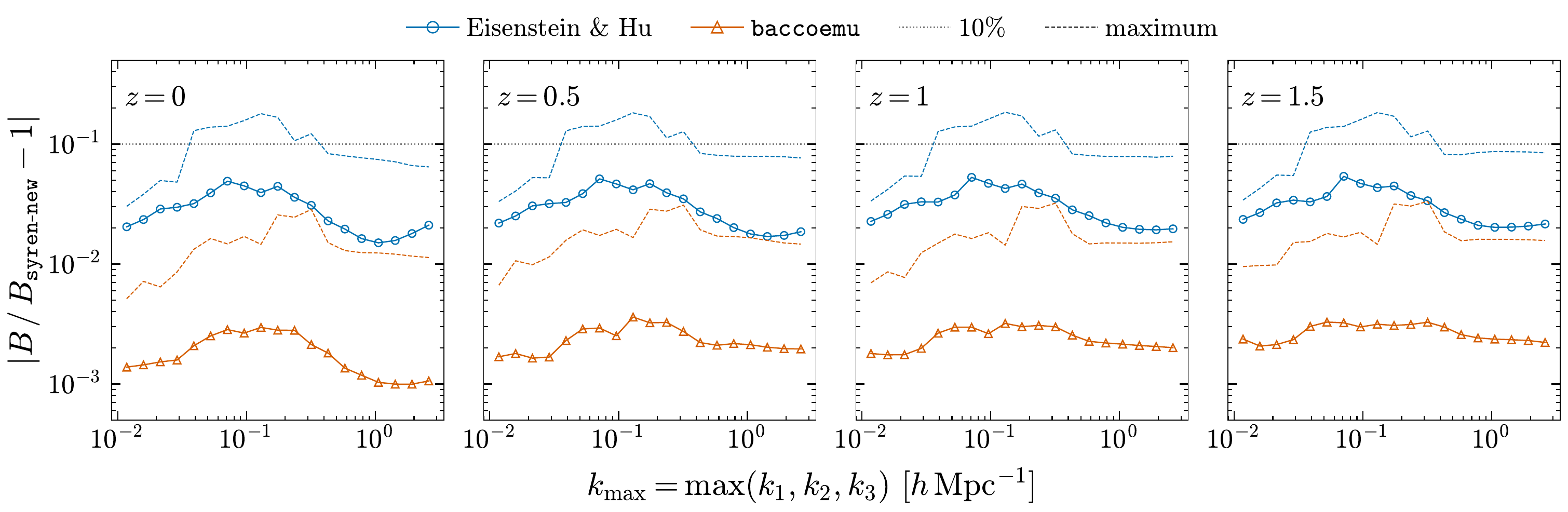}
\caption{Mean (solid) and maximum (dashed) fractional differences of the matter bispectrum for different linear power spectrum backends, compared to the \texttt{syren-new} backend. We show this against $k_{\rm max}=\max(k_1,k_2,k_3)$ for \texttt{bihalofit-JAX} with the Eisenstein \& Hu (E\&H) and \texttt{baccoemu} linear power spectra, over $1000$ cosmologies and the triangles in each $k_{\rm max}$ bin, at four redshifts. The triangles are those of Figure~\ref{fig:matter_consistency}, the cosmologies are those of the benchmark set (Section~\ref{sec:implementations}) and all runs use \texttt{float32}, the precision of the \texttt{baccoemu} network.}
\label{fig:backend_consistency}
\end{figure*}

\begin{figure*}[htbp]
\centering
\includegraphics[width=\textwidth]{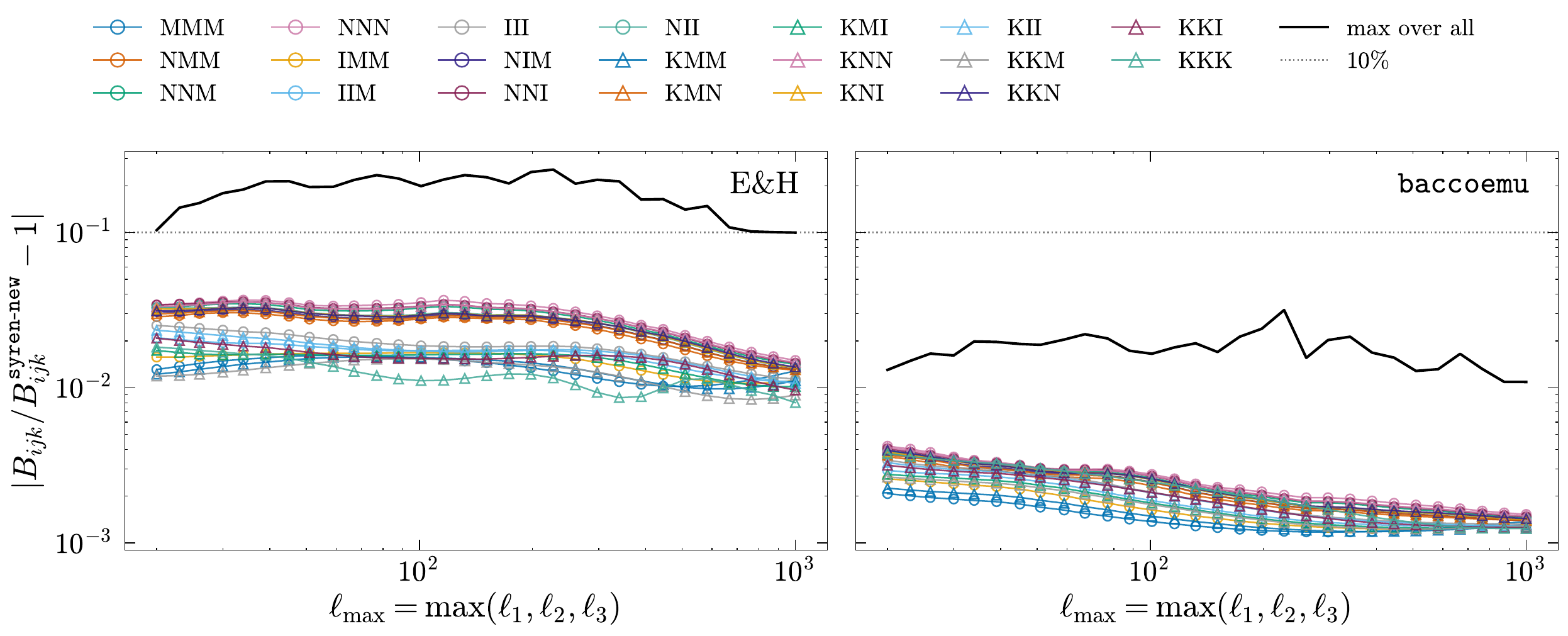}
\caption{Mean difference of the projected bispectrum $B_{ijk}$ for \texttt{bihalofit-JAX} in \texttt{float32} with the Eisenstein \& Hu (left) and \texttt{baccoemu} (right) linear power spectrum, relative to \texttt{syren-new}, averaged over the $200$ cosmologies of the convergence tests and the triangles with each $\ell_{\rm max}=\max(\ell_1,\ell_2,\ell_3)$, for the twenty tracer combinations (M lensing, N counts, I intrinsic alignment, K CMB lensing) at the default $N_\chi=160$. The black line is the maximum over every combination, cosmology and triangle.}
\label{fig:backend_bkappa}
\end{figure*}

The linear power spectrum (Section~\ref{sec:backends}) enters every statistic considered in this work, so it is important to understand what effect any mismodelling there might introduce. To assess this we run \texttt{bihalofit-JAX} at its default settings in \texttt{float32} ($N_R=15$, $N_k=600$), the precision of the \texttt{baccoemu} network, with each backend, and compare E\&H and \texttt{baccoemu} with \texttt{syren-new}. For the matter bispectrum (Figure~\ref{fig:backend_consistency}), \texttt{baccoemu} differs from \texttt{syren-new} by a mean of $0.2$--$0.3\%$ (at most $3.3\%$) and the Eisenstein \& Hu spectrum by a mean of $3\%$ (at most $18\%$), with little dependence on $k_{\rm max}$ or redshift. For the projected bispectrum (Figure~\ref{fig:backend_bkappa}) the means are $0.13$--$0.22\%$ and $1.1$--$2.8\%$, largest for the combinations with counts legs (up to $3.2\%$ and $25\%$ at the worst triangle). The skew spectra and aperture statistics are weighted sums over triangles of the same bispectrum, so for the same tracers their differences are expected to be bounded by those of $B_{ijk}$; this is what we find for the combinations of the convergence tests, with means of $0.1$--$0.2\%$ for \texttt{baccoemu} and $1.1$--$2.5\%$ for E\&H, and maxima below $1.5\%$ and $13\%$. These differences isolate the effect of the linear power spectrum and say nothing about the accuracy of BiHalofit against simulations.

\subsection{Performance}

The methodology of timing here follows the same procedure that was described in Section~\ref{sec:benchmarks}. Table~\ref{tab:backend_timings} gives the wall time of \texttt{bihalofit-JAX} with each backend, for the workloads of Section~\ref{sec:benchmarks}. \texttt{baccoemu} costs within $20\%$ of \texttt{syren-new}. The E\&H backend is slower on the GPU, by about $2\,\text{ms}$ in \texttt{float32} and $5$--$7\,\text{ms}$ in \texttt{float64} per call, and within $35\%$ on the CPU. The extra time is the solution of the linear growth ODE, which the other backends avoid: timed alone on the GPU it takes $2$--$8\,\text{ms}$, against under $0.5\,\text{ms}$ for a whole call of the other two.

\citet{Sui_2025} find that \texttt{syren-new} is $\sim 8$ and $\sim 10$ times faster than \texttt{baccoemu} on a CPU and a GPU (at $200$ wavenumbers, against the $600$ that \texttt{baccoemu} always computes), whereas here the two cost the same, since the difference in the cost of their linear power spectra is subdominant in the full calculation. Their CPU times are averaged over calls of one cosmology each and their GPU times are per cosmology in batches of $2000$, so most of their difference between CPU and GPU comes from batching; for one cosmology the fixed cost of a GPU call dominates, and we find the CPU faster. 

\begin{table*}[htbp]
\centering
\small
\setlength{\tabcolsep}{3pt}
\begin{tabular}{l l c r r r r r r r r r r}
\hline\hline
 & & & \multicolumn{4}{c}{\texttt{syren-new} [ms]} & \multicolumn{4}{c}{E\&H [ms]} & \multicolumn{2}{c}{\texttt{baccoemu} [ms]} \\
\cline{4-7}\cline{8-11}\cline{12-13}
 & & & \multicolumn{2}{c}{GPU} & \multicolumn{2}{c}{CPU} & \multicolumn{2}{c}{GPU} & \multicolumn{2}{c}{CPU} & GPU & CPU \\
\cline{4-5}\cline{6-7}\cline{8-9}\cline{10-11}\cline{12-12}\cline{13-13}
Stat. & Workload & $N_{\rm cosmo}$ & \texttt{fp32} & \texttt{fp64} & \texttt{fp32} & \texttt{fp64} & \texttt{fp32} & \texttt{fp64} & \texttt{fp32} & \texttt{fp64} & \texttt{fp32} & \texttt{fp32} \\
\hline
$B_\delta$ & $10{,}160$ & $1$ & $1.44$ & $1.52$ & $0.932$ & $1.41$ & $3.6$ & $8.35$ & $0.859$ & $1.05$ & $1.55$ & $0.832$ \\
 & (triangles) & $500$ & $1.75$ & $15.3$ & $62.7$ & $146$ & $4.45$ & $22.2$ & $61.3$ & $139$ & $1.9$ & $57.1$ \\
$B_\gamma$ & $9{,}799$ & $1$ & $2.48$ & $9.11$ & $25.3$ & $43.7$ & $4.57$ & $15.5$ & $23.5$ & $49.8$ & $2.51$ & $26.4$ \\
 & (triangles) & $500$ & $83.4^\ast$ & $2{,}741^\ast$ & $8{,}514$ & $17{,}060$ & $87.1^\ast$ & $2{,}758^\ast$ & $8{,}230$ & $17{,}015$ & $86.2^\ast$ & $8{,}090$ \\
$C_\ell^{g^2g}$ & $39$ & $1$ & $7.66$ & $163$ & $10{,}125$ & $9{,}870$ & $11$ & $168$ & $8{,}908$ & $9{,}296$ & $7.81$ & $9{,}182$ \\
 & (multipoles) & $500$ & $1{,}399^\ast$ & $73{,}913^\ast$ & $263{,}036$ & $507{,}975$ & $1{,}377^\ast$ & $74{,}024^\ast$ & $237{,}169$ & $477{,}948$ & $1{,}376^\ast$ & $240{,}367$ \\
$\langle M_{\rm ap}^3\rangle$ & $84$ & $1$ & $5.95$ & $252$ & $1{,}059$ & $1{,}506$ & $8.68$ & $257$ & $1{,}417$ & $1{,}499$ & $5.96$ & $1{,}125$ \\
 & (triplets) & $500$ & $2{,}113^\ast$ & $143{,}529^\ast$ & $400{,}378$ & $853{,}978$ & $2{,}061^\ast$ & $150{,}572^\ast$ & $359{,}205$ & $810{,}050$ & $2{,}082^\ast$ & $367{,}053$ \\
\hline
\end{tabular}

\refstepcounter{table}
\label{tab:backend_timings}
\begin{minipage}{\textwidth}
\footnotesize
\justifycaption\noindent\textbf{Table \thetable.} Wall time of \texttt{bihalofit-JAX} for each linear power spectrum backend, measured as in Table~\ref{tab:projected_speeds} (a CPU restricted to $24$ cores). \texttt{baccoemu} is listed in \texttt{float32} only, since its network evaluates in \texttt{float32}. $^\ast$Runs in the sequential batches of Table~\ref{tab:projected_speeds}. The \texttt{syren-new} columns are those of Table~\ref{tab:projected_speeds}; each entry is the median over $1$--$10$ processes.
\end{minipage}
\end{table*}

\bibliographystyle{aasjournal}
\bibliography{paper}

\end{document}